\documentclass[twocolumn, twocolappendix]{aastex631}

\usepackage{amsmath, mathtools, amsthm, amssymb}
\usepackage{enumitem}
\usepackage[toc,page]{appendix}
\hypersetup{colorlinks=true, allcolors=cyan}
\usepackage{soul}
\usepackage{graphicx}
\usepackage{appendix}
\graphicspath{ {./} }
\usepackage{xcolor}
\usepackage{hyperref}

\newcommand{\mstellar}{\ensuremath{M_{\star}}}

\newcommand{\msol}{\ensuremath{\rm{M_{\odot}}}}
\newcommand{\re}{\ensuremath{{\rm R_{50}}}}

\newcommand{\NH}{\mbox{\sc \small NewHorizon}}

\usepackage[super]{nth}

\shorttitle{Lenticular galaxy formation in fields}

\shortauthors{Han et al.}

\begin{document}

\title{Exploring lenticular galaxy formation in field environments using {\sc \small NewHorizon}: evidence for counter-rotating gas accretion as a formation channel}

\email{E-mail: genesis11@yonsei.ac.kr, yi@yonsei.ac.kr}

\author{\href{https://orcid.org/0009-0004-4349-006X}{Seongbong Han} \href{https://orcid.org/0009-0004-4349-006X}{\includegraphics[scale=0.5]{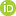}}}
\affil{Department of Astronomy and Yonsei University Observatory, 50 Yonsei-ro, Seodaemun-gu, 
Seoul 03722, Republic of Korea}

\author{\href{https://orcid.org/0000-0002-0858-5264}{J. K. Jang} \href{https://orcid.org/0000-0002-0858-5264}{\includegraphics[scale=0.5]{ORCIDiD_icon16x16.png}}}
\affil{Department of Astronomy and Yonsei University Observatory, 50 Yonsei-ro, Seodaemun-gu, 
Seoul 03722, Republic of Korea}

\author{\href{https://orcid.org/0000-0002-2873-8598}{Emanuele Contini} \href{https://orcid.org/0000-0002-2873-8598}{\includegraphics[scale=0.5]{ORCIDiD_icon16x16.png}}}
\affil{Department of Astronomy and Yonsei University Observatory, 50 Yonsei-ro, Seodaemun-gu, 
Seoul 03722, Republic of Korea}

\author{\href{https://orcid.org/0000-0003-0225-6387}{Yohan Dubois} \href{https://orcid.org/0000-0003-0225-6387}{\includegraphics[scale=0.5]{ORCIDiD_icon16x16.png}}}
\affil{Institut d’Astrophysique de Paris, CNRS, Sorbonne Université, UMR 7095, 98 bis bd Arago, 75014 Paris, France}

\author{\href{https://orcid.org/0000-0002-1270-4465}{Seyong Jeon} \href{https://orcid.org/0000-0002-1270-4465}{\includegraphics[scale=0.5]{ORCIDiD_icon16x16.png}}}
\affil{Department of Astronomy and Yonsei University Observatory, 50 Yonsei-ro, Seodaemun-gu, 
Seoul 03722, Republic of Korea}

\author{\href{https://orcid.org/0000-0002-5601-575X}{Sugata Kaviraj} \href{https://orcid.org/0000-0002-5601-575X}{\includegraphics[scale=0.5]{ORCIDiD_icon16x16.png}}}
\affil{Centre for Astrophysics Research, University of Hertfordshire, Hatfield AL10 9AB, UK}

\author{\href{https://orcid.org/0000-0002-3950-3997}{Taysun Kimm} \href{https://orcid.org/0000-0002-3950-3997}{\includegraphics[scale=0.5]{ORCIDiD_icon16x16.png}}}
\affil{Department of Astronomy and Yonsei University Observatory, 50 Yonsei-ro, Seodaemun-gu, 
Seoul 03722, Republic of Korea}

\author{\href{https://orcid.org/0000-0001-6180-0245}{Katarina Kraljic} \href{https://orcid.org/0000-0001-6180-0245}{\includegraphics[scale=0.5]{ORCIDiD_icon16x16.png}}}
\affil{Observatoire Astronomique de Strasbourg, Université de Strasbourg, CNRS, UMR 7550, F-67000 Strasbourg, France}

\author{\href{https://orcid.org/0000-0002-4731-9604}{Sree Oh} \href{https://orcid.org/0000-0002-4731-9604}{\includegraphics[scale=0.5]{ORCIDiD_icon16x16.png}}}
\affil{Department of Astronomy and Yonsei University Observatory, 50 Yonsei-ro, Seodaemun-gu, 
Seoul 03722, Republic of Korea}

\author{\href{https://orcid.org/0000-0001-6902-2898}{Sébastien Peirani} \href{https://orcid.org/0000-0001-6902-2898}{\includegraphics[scale=0.5]{ORCIDiD_icon16x16.png}}}
\affil{Institut d’Astrophysique de Paris, CNRS, Sorbonne Université, UMR 7095, 98 bis bd Arago, 75014 Paris, France}
\affil{Université Côte d’Azur, Observatoire de la Côte d’Azur, CNRS, Laboratoire Lagrange, Bd de l’Observatoire, CS 34229, 06304 Nice Cedex 4, France}
\affil{Department of Physics, School of Science, The University of Tokyo, 7-3-1 Hongo, Bunkyo-ku, Tokyo 113-0033, Japan
}

\author{\href{https://orcid.org/0000-0003-0695-6735}{Christophe Pichon} \href{https://orcid.org/0000-0003-0695-6735}{\includegraphics[scale=0.5]{ORCIDiD_icon16x16.png}}}
\affil{Institut d’Astrophysique de Paris, CNRS, Sorbonne Université, UMR 7095, 98 bis bd Arago, 75014 Paris, France}
\affil{Université Paris-Saclay, CNRS, CEA, Institut de physique théorique, 91191 Gif-sur-Yvette, France}
\affil{Korea Institute for Advanced Studies (KIAS), 85 Hoegi-ro, Dongdaemun-gu, Seoul 02455, Republic of Korea}

\author{\href{https://orcid.org/0000-0002-4556-2619}{Sukyoung K. Yi} \href{https://orcid.org/0000-0002-4556-2619}{\includegraphics[scale=0.5]{ORCIDiD_icon16x16.png}}}
\affil{Department of Astronomy and Yonsei University Observatory, 50 Yonsei-ro, Seodaemun-gu, 
Seoul 03722, Republic of Korea}



\begin{abstract}
The formation pathways of lenticular galaxies (S0s) in field environments remain a matter of debate. We utilize the cosmological hydrodynamic simulation, \NH, to investigate the issue. We select two massive star-formation quenched S0s as our main sample. By closely tracing their physical and morphological evolution, we identify two primary formation channels: mergers and counter-rotating gas accretion. The former induces central gas inflow due to gravitational and hydrodynamic torques, triggering active central star formation which quickly depletes the gas of the galaxy. Counter-rotating gas accretion overall has a similar outcome but more exclusively through hydrodynamic collisions between the pre-existing and newly-accreted gas. Both channels lead to S0 morphology, with gas angular momentum cancellation being a crucial mechanism. These formation pathways quench star formation on a short timescale ($<$\,Gyr) compared to the timescales of environmental effects. We also discuss how counter-rotating gas accretion may explain the origin of S0s with ongoing star formation and the frequently observed gas-star misaligned kinematics in S0s.
\end{abstract}

%
\section[]{Introduction}
\label{sec: introduction}

The class of S0 galaxies was originally introduced to bridge the morphological transition between elliptical and spiral galaxies in the Hubble's tuning fork scheme \citep[][]{hubble1936}. S0s were traditionally characterized by a flattened shape without spiral arms, a prominent bulge, and red colors. Research on their formation mechanisms has been strongly affected by the morphology-density relation \citep[][]{dressler1980morphdensity, goto2003morphdensity, postman2005morphdensity}, which reveals a higher proportion of S0s in high-density environments at the expense of a decreasing fraction of late-type galaxies. Various scenarios have been proposed to explain this relation, including ram pressure stripping of gas from late-type galaxies as they pass through the hot intra-cluster medium (ICM) \citep[][]{gunn1972rampressure, quilis2000rampressure}, and the removal of halo gas via hydrodynamical interaction with ICM that prevents further cold gas accretion and shuts down star formation \citep[][]{larson1980strangulation, bekki2002strangulation}.

However, the high fraction of S0 galaxies in group or subgroup environments at low redshifts \citep[][]{wilman2009fieldS0, just2010S0frac}, along with the difficulties in interpreting their scaling relation solely by faded spiral scenarios \citep[e.g.,][]{bedregal2006S0TFR,williams2010TFR}, suggests that previous models may not fully account for S0 formation. Numerical studies have shown that both minor and major mergers are able to produce S0 remnants, affecting stellar kinematics and structures \citep[][]{querejeta2015merger, eliche2018merger}. In contrast, \cite{rizzo2018fadedspiral} argued that mergers are not primary pathways in low-density environments. Their analysis of stellar kinematics from CALIFA (Calar Alto Legacy Integral Field Area) \citep[][]{sanchez2012CALIFA, garcia2015CALIFA} survey data revealed that most S0s follow the trend of spiral galaxies on the stellar specific angular momentum versus stellar mass ($j_{\star}-M_{\star}$) plane. They proposed that spirals can transform to S0s through a passive evolution driven by active galactic nuclei (AGN) feedback \citep[][]{van2009lenticular} or the cessation of cold gas supply \citep[e.g.,][]{dekel2006shutdown, peng2015strangulation, armillotta2016shutdown}. Recent SAMI (Sydney-AAO Multi-object Integral-field Spectrograph) \citep[][]{croom2012SAMI, bryant2015SAMI} survey findings reveal that many S0 galaxies in low-mass group and field environments exhibit weaker stellar rotation and signs of disruptive merger events, such as shells and tidal streams, suggesting their merger origins \citep[][]{deeley2020SAMI_S0}.

Since \cite{galletta1987counter} first observed the gas-star counter-rotation in a barred S0 galaxy, numerous similar cases have been documented \citep[][]{bertola1992counter, rubin1992counter, kuijken1996counter}. Studies indicate that ellipticals and S0s more frequently harbor counter-rotating gaseous components compared to spiral galaxies \citep[][]{kannappan2001counter, pizzella2004counter, bureau2006counterrotatinggas, katkov2015counter}. These findings have spurred further investigations into S0 formation pathways involving kinematic misalignment.

\cite{bassett2017retrograde} conducted hydrodynamical simulations of minor mergers between a disc galaxy and a gas-rich satellite across various merger parameters, discovering that a retrograde orbit is essential for generating gas-star counter-rotation in an S0 galaxy. They also noted that a large amount of gas in the primary galaxy can hinder the formation of a counter-rotating gas disk. \cite{osman2017counterrotatinggasquenching} proposed that a high fraction of counter-rotating gas disturbs spiral arm formation. Using the sample of 401 emission-line S0s from the MaNGA (Mapping Nearby Galaxies at Apache Point Observatory) \citep[][]{blanton2017MANGA} survey data, \cite{zhou2023misalinged} found that $\sim20\%$ of S0s exhibit gas-star misalignment, highlighting misaligned gas accretion as an important channel for S0 formation.

A recent study on S0 formation using a large-scale cosmological simulation was conducted by \cite{deeley2021S0TNG}. They utilized the IllustrisTNG100 simulation \citep{pillepich2018TNG} to explore S0 formation across a range of environmental settings, from fields to high-mass groups. They identified merger events and gas stripping due to group infall as the major formation pathways, with a smaller contribution from passive evolution through gas exhaustion. However, S0s with counter-rotating components in the simulation were not discussed.

In this work, we aim to study S0 formation in low-density environments using the \NH\ simulation \citep[][]{dubois2021introducing}, a cosmological hydrodynamic zoom-in simulation. \NH\ is particularly well-suited to our study due to its high time output cadence and its excellent mass and spatial resolutions, which enable us to trace the detailed evolution of physical properties and galactic morphology. Furthermore, it predominantly covers field and low-mass group environments, aligning with our primary research interests.

We demonstrate in this study that S0 formation can be driven by both mergers and counter-rotating gas accretion. In both scenarios, a decrease in gas angular momentum plays a key role in building S0 morphology by relocating exterior gas to the galactic center.

In Section~\ref{sec: Methods}, we briefly describe the \NH\ simulation. We then detail the sample selection process and the analysis. In Section~\ref{sec: S0 formation pathways}, we examine the evolution of S0 galaxies’ physical properties and morphology to identify key factors determining their final morphology. We further discuss the formation pathways identified in this study and their physical mechanisms in Section~\ref{sec: Discussion}. Finally, in Section~\ref{sec: Summary and conclusions}, we summarize our findings and highlight their implications.

%
\section[]{Methods}
\label{sec: Methods}

\subsection{\NH}
\label{sec: Methods - NEWHORIZON}

\NH\ (NH) has run with the adaptive mesh refinement (AMR) code RAMSES \citep[][]{teyssier2002cosmological}, covering a spherical volume with a 10\,cMpc radius within the HORIZON-AGN simulation \citep[][]{dubois2016thehorizon}. The AMR spatial resolution reaches 34\,pc, and the mass resolutions of dark matter (DM) and stellar particles are $10^6$ and $10^4 \,\msol$, respectively. NH is implemented with cosmological parameters compatible with the WMAP-7 $\Lambda$CDM cosmology: $\Omega_{\rm m}$ = 0.272, $\Omega_{\Lambda}$ = 0.728, $\sigma_8$ = 0.81, $\Omega_{\rm b}$ = 0.045, $H_{0}$ = 70.4 $\rm km \, \rm s^{-1} \rm Mpc^{-1}$, and $n_{\rm s}$ = 0.967 \citep[][]{komatsu2011seven}. Down to redshift $z=0.17$, the simulation has produced 863 snapshots with a time interval of $\sim15$\,Myr.

As the specifics of the simulation are detailed in \cite{dubois2021introducing}, we briefly present here the key sub-grid physics. Stellar particles form within cells where the hydrogen gas number density exceeds $n_{\rm H} = 10\,\rm cm^{-3}$, following the Kennicutt-Schmidt relation \citep[][]{schmidt1959therate, kennicutt1998theglobal}. For those cells, the star formation efficiency is determined by the hydrodynamic properties, such as the turbulent Mach number \citep[][]{kimm2017feedbackregulated}. Regarding stellar feedback, type II SN feedback is implemented, assuming a Chabrier initial mass function (IMF) \citep[][]{chabrier2003IMF} for stellar particles. SNe explode 5\,Myr after the formation of a stellar particle, releasing kinetic energy of $10^{51}\,\rm erg$.
Considering the IMF and SN rates, a stellar particle returns $\sim31\%$ of its initial mass to neighboring gas cells. Black holes (BHs), modeled as sink particles, can release two types of AGN feedback depending on the ratio of their accretion rate to their Eddington rate \citep[][]{dubois2012AGN}: radio and quasar modes. BHs release energy in the form of mass, momentum, and total energy by the jet mode \citep[][]{dubois2010jet}, while in the quasar mode, they release only thermal energy \citep[][]{teyssier2011quasar}.

\subsection{Sample selection}
\label{sec: Methods - Sample selection}

\subsubsection{Halo and galaxy detection}
\label{sec: Methods - Sample selection - Halo and galaxy detection}

We used the AdaptaHop \citep[][]{aubert2004adaptahop, tweed2009} algorithm to identify dark matter haloes and galaxies. The virial radius and mass of haloes ($R_{\rm vir}$ and $M_{\rm vir}$) are calculated based on the energy criterion of virialization. We identify three low-mass group-sized haloes ($10^{12.5} < \textit M_{\rm vir}/\msol < 10^{13}$) at $z=0.17$.

Galaxies are initially identified if they contain at least 100 stellar particles. We detect 214 galaxies with the total mass of stellar particles greater than $10^9\,\msol$ at $z=0.17$. The galaxy merger tree is constructed using the membership of stellar particles in galaxies across snapshots.

\subsubsection{Target S0 galaxies}
\label{sec: Methods - Sample selection - Target S0 galaxies}

What defines an S0 galaxy? Although there is no universal agreement on the precise features that characterize an S0, common attributes include a visible stellar disk, often accompanied by an envelope, a dust lane, or a prominent bulge \citep[e.g.,][]{sandage1961atlas, baillard2011S0morph}. However, the broad characteristics of S0 galaxies, overlapping with those of other types, reflect their diverse evolutionary histories, complicating the establishment of a clear-cut definition \citep[e.g.,][]{laurikainen2010BTofS0, dom2020kin_SP_S0, deeley2020SAMI_S0}. In this regard, we avoid using complicated morphological constraints or parametric definitions for S0 selection, as they may limit our understanding of S0s by overlooking the evolutionary diversity. Instead, we rely on visual classification, focusing on a basic criterion for S0 morphology, \textit{a visible stellar disk lacking spiral arms}.

Galactic morphology is broadly linked to star formation activities \citep[][]{robert1994morphSF, blanton2003morphSF, kinyumu2024morphSF}. Massive early-type galaxies (ETGs) typically belong to the red-sequence \citep[][]{schawinski2009earlytyperedsequence}, although the trend is unclear in the dwarf regime \citep[][]{lazar2024dwarfETG}. However, recent or ongoing star formation in ETGs challenges the traditional view of them as `red and dead' \citep[e.g.,][]{yi2005starformingE, temi2009starformingS0}, implying complexity in their evolutionary process. For instance, the question remains whether S0s with notable star-forming activities are simply passively evolving spirals or if they are driven by external factors, such as minor mergers \citep[e.g.,][]{xu2022SFS0}. To perform a comprehensive study on S0 galaxies but with a streamlined organization, we first study \textit{quenched S0s} as our main targets in Section~\ref{sec: S0 formation pathways} while also discussing the possible origins of \textit{star-forming S0s} in Section~\ref{sec: Discussion}.

\subsubsection{Mock image generation}
\label{sec: Methods - Sample selection - Mock image generation}

For visual classification and investigation in this study, we produce mock observation images of NH galaxies. Our mock images are generated with SKIRT~\citep{baes2015SKIRT, camps2020SKIRT9}, a three-dimensional Monte Carlo radiative transfer code, which simulates the effect of dust on photons emitted from light sources, accounting for absorption, emission, and scattering. We apply a simple stellar population model of \cite{bruzual2003stellar} with a Chabrier IMF to each star particle. For gas cells with temperatures below 30,000\,K, we assign the THEMIS dust population model \citep{jones2016THEMIS}. We set the dust-to-metal ratio to 0.4 and use 8 bins for the grain size distributions of silicate, graphite, and polycyclic aromatic hydrocarbon dust. We set the number of photon packets to $2\times10^7$. The pixel size is 40\,pc, which is comparable to the spatial resolution of the simulation. The images are created by combining red, green, and blue colors corresponding to the i-, r-, and g-band fluxes of Sloan Digital Sky Survey (SDSS) \citep{york2000SDSS} in the rest frame.

\subsubsection{Sampling process}
\label{sec: Methods - Sample selection - Sampling process}

The stages of our sampling process are as follows: (i) We extract galaxies with stellar masses (\mstellar) ranging from $10^{10}$ to $10^{11}\,\msol$ at $z=0.17$. Here, \mstellar~is defined as the total mass of stellar particles inside $2\re$, where \re~is the radius of a sphere containing half the mass of all stellar particles of a galaxy\footnote{We use 2\re\ as the boundary of a galaxy in this study to consider the difficulty in measuring the properties of bound particles or stars in the simulation and observations.}. We identified 40 galaxies within this mass range. (ii) For visual classification, we generate mock images of these galaxies viewed from 12 angles, comprising 3 azimuthal angles (0$^{\circ}$, 90$^{\circ}$, and 180$^{\circ}$) and 4 inclination angles (0$^{\circ}$, 30$^{\circ}$, 60$^{\circ}$, and 90$^{\circ}$) for their face-on views. (iii) These mock images were then visually inspected by 8 researchers, who classified the galaxies into five types: elliptical (E), S0, late/S0, late-type, and irregular (Irr). For S0 classification, the focus was on whether a galaxy has a visible stellar disk and lacks spiral arms. In order to exclude controversial cases, we introduce the late/S0 type, which represents disk galaxies that are difficult to classify as either S0 or late-type due to an insignificant spiral pattern or ambiguous shape. The morphological type of a galaxy is determined based on the consensus of more than half of the inspectors. If consensus is not reached, the galaxy is classified as late/S0 if more than half recognize it as a disk galaxy (S0, late/S0, or late) or as Irr otherwise. (iv) Finally, to distinguish between quenched and star-forming galaxies, we use the definition of quenched galaxies from \cite{weinberger2018SFMS}. They use the star-forming main sequence (SFMS) relation derived by \cite{ellison2015SFMS_M}:
\begin{equation}
\begin{array}{cl}
    \log(\frac{{\rm SFR}_{\rm SFMS}}{\msol{\rm/yr}}) = -7.4485 + 0.7575 \log(\frac{\mstellar}{\msol}) \\ + 1.5 \log(1+z), \end{array}
    \label{eq1_SFMS}
\end{equation}

where the redshift dependence is adopted from \cite{schreiber2015SFMS_z} and logarithm denotes the decimal logarithm. According to their definition, a galaxy is considered quenched if its SFR\footnote{SFR is defined by the total mass of stellar particles younger than 200\,Myr within 2\re~divided by 200\,Myr, the star formation timescale which near-ultraviolet emission traces \citep[][]{kennicutt2012SFR}. We use this to avoid temporal fluctuations that may be caused by using shorter timescales.} is 1\,dex below the SFMS for their corresponding \mstellar~and redshift. As mentioned, we select quenched S0 galaxies as our main targets.

\begin{figure}
\centering
\includegraphics[width=0.47\textwidth]{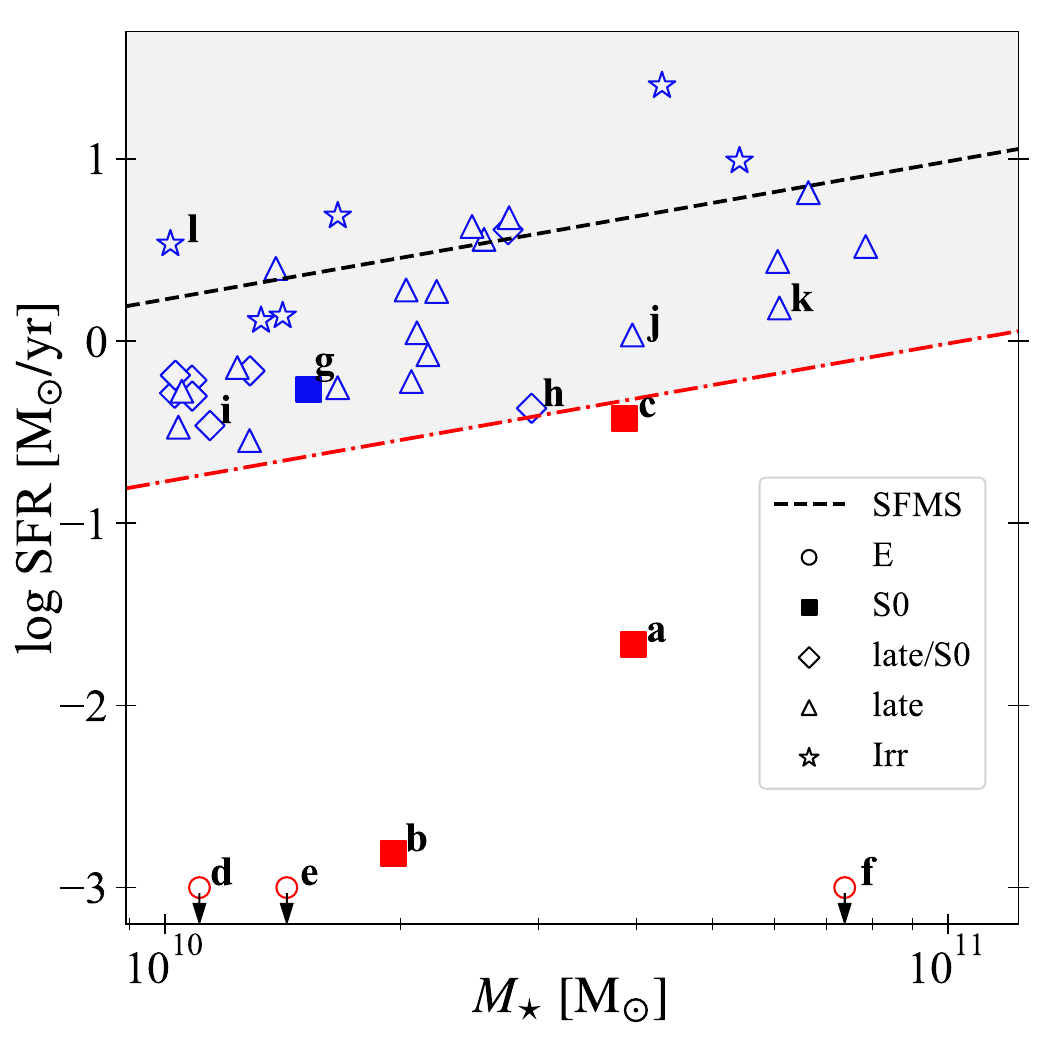}
\caption{SFR as a function of \mstellar~for NH galaxies with $10^{10}<\textit\mstellar/\msol<10^{11}$ at $z=0.17$.
The morphological types of galaxies are indicated with different marker shapes.
Galaxies with an SFR below the red dotted line are considered quenched, which is 1\,dex below the SFMS (black dashed line).
Red markers denote quenched galaxies, while blue markers indicate star-forming galaxies.
The labeled markers correspond to the twelve galaxies with mock observation images displayed in Figure~\ref{fig02_12gals_mockimgs}.
SFRs lower than $10^{-3}\,\msol\rm/yr$ are plotted at this value, applicable only to Es.
 }
\label{fig01_SFR_M}
\end{figure}

Figure~\ref{fig01_SFR_M} presents the result of our visual morphology classification along with the SFRs and \mstellar~of NH galaxies. The black dashed line indicates the SFMS, and the red dotted line represents $\rm SFR_{\rm SFMS}$$\,-\,$1\,dex, below which galaxies are considered quenched. Three galaxies (a, b, and c) are S0s with quenched star formation, meeting the criteria to be our primary targets. All Es (d, e, and f) exhibit substantially low SFRs, less than $10^{-3}\,\msol/yr$. We identify one star-forming S0 (g).

\begin{figure*}
\centering
\includegraphics[width=0.9\textwidth]{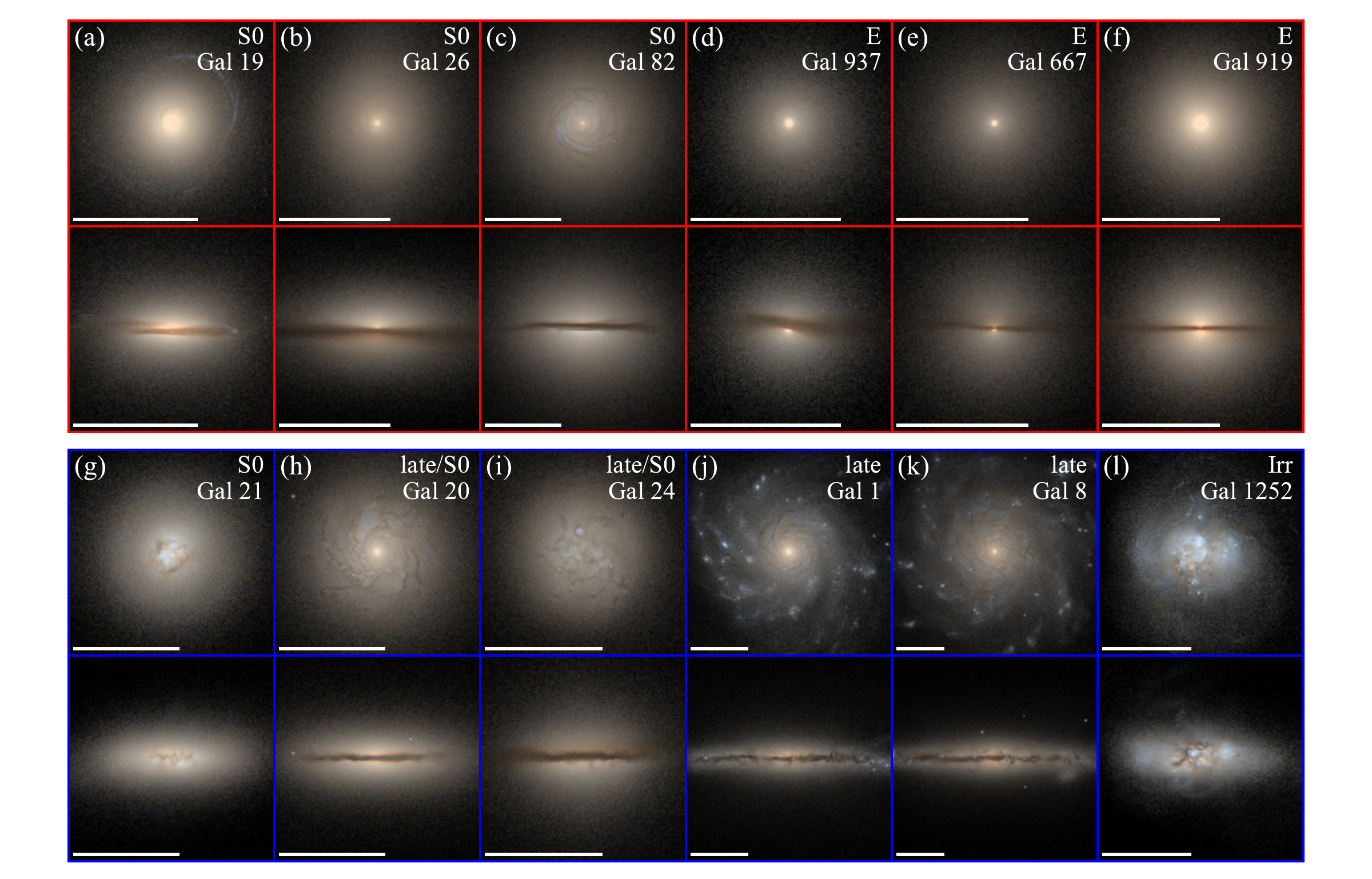}
\caption{Mock observation images of the twelve galaxies labeled in Figure~\ref{fig01_SFR_M}, shown in both face-on and edge-on views at $z=0.17$.
Panels for quenched galaxies and star-forming galaxies are framed in red and blue lines, respectively. 
The morphological type and ID of each galaxy are indicated in the face-on view panels.
The white bars indicate 10\,kpc.
 }
\label{fig02_12gals_mockimgs}
\end{figure*}

Figure~\ref{fig02_12gals_mockimgs} shows mock images of the twelve galaxies labeled in Figure~\ref{fig01_SFR_M} from both face-on and edge-on views. In Appendix~\ref{sec: Appendix - Morphology verification}, we verify that the visual stellar disks are reasonably detected and provide the mock images of galaxies not shown in Figure~\ref{fig02_12gals_mockimgs}. For simplicity, we henceforth refer to galaxies by their morphological type and ID number (e.g., S0\,19).

\subsubsection{Further examination}
\label{sec: Methods - Sample selection - Further examination}

Unfortunately, however, S0\,82 is heavily contaminated by low-resolution DM particles coming from the outer region of NH, making up $\sim98\%$ of the total DM mass within its 2\re. This may increase errors in calculating gravitation between grids, possibly bringing about unphysical effects \citep[e.g.,][]{onorbe2014contam}. Thus, we do not use S0\,82 in this study.

\begin{figure}
\centering
\includegraphics[width=0.5\textwidth]{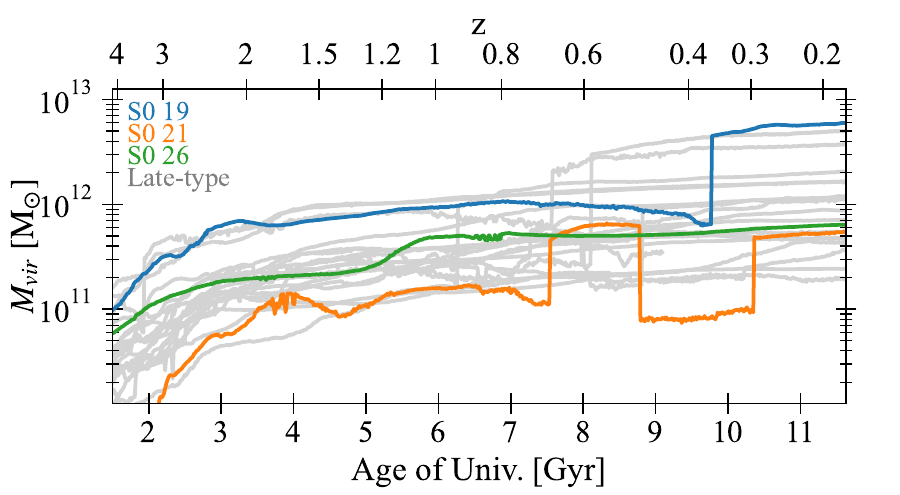}
\caption{Evolution of the $M_{\rm vir}$ of halos in which NH galaxies reside.
Sudden jumps in $M_{\rm vir}$ represent the infall of the galaxy into a more massive halo.
The histories of S0s are shown in different colors.
For comparison, those of all late-type galaxies are shown in gray.
 }
\label{fig03_env}
\end{figure}

In order to ensure that dense environments do not influence the formation of NH S0s, we examine their environmental history. Figure~\ref{fig03_env} shows $M_{\rm vir}$ of the most massive halo where galaxies reside within $R_{\rm vir}$ from the halo center as a function of time. Therefore, $M_{\rm vir}$ steeply increases when the galaxy enters a more massive halo than before. S0\,26 remains a central galaxy within the halo that does not grow to group size ($M_{\rm vir} > 10^{12.5}\,\msol$) throughout its entire history. Although S0\,19 enters a group-sized halo at $z \sim 0.4$, this occurs only after it develops its S0 morphology, as will be detailed in the next section. S0\,21 exhibits orbital motion at $z < 0.8$ around a low-mass halo ($M_{\rm vir}\sim 10^{11.5}\,\msol$). In Section~\ref{sec: Discussion - Star-forming S0}, we discuss that its transformation into an S0 is attributed to counter-rotating gas accretion (since $z \sim 0.4$) rather than environmental effects.

\subsection{Quenching time scale}
\label{sec: Methods - Quenching time scale}

As outlined in Section~\ref{sec: Methods - Sample selection - Sampling process}, we define quenched galaxies as those with an SFR 1\,dex below the SFMS for a given redshift. 
We denote $t_{\rm Q}$ as the time when a galaxy becomes quenched. These times are measured with respect to the age of the universe. 
The beginning of quenching ($t_{\rm BQ}$) is defined as the time when the SFR first drops to 0.5\,dex below the SFMS, looking back from $t_{\rm Q}$. 
To quantify the duration of the quenching process, we use the quenching time scale ($\tau_{\rm Q}$), which is defined as the time interval between $t_{\rm Q}$ and $t_{\rm BQ}$. For our main targets, the quenched S0s, their SFRs do not fluctuate significantly, so measuring $\tau_{\rm Q}$ is straightforward.

\subsection{Merger identification}
\label{sec: Methods - Merger identification}

For merger identification, we take advantage of our galaxy merger tree. The occurrence time of a merger event is defined as the moment when the total mass of stellar particles of a companion galaxy ($M_{\rm \star,c}$) first drops below 50$\%$ of its maximum value in its history. If the last progenitor of the companion ends before this mass threshold is reached, the merger time is set to the time of the last progenitor. We categorize mergers into three types based on the stellar mass ratio between a companion and a main galaxy ($M_{\rm \star,c}/M_{\rm \star,m}$), which is measured when $M_{\rm \star,c}$ reaches its maximum (i.e., when the companion begins to lose its stellar mass): major mergers ($M_{\rm \star,c}/M_{\rm \star,m}$ $>$ 0.25), minor mergers (0.05 $<$ $M_{\rm \star,c}/M_{\rm \star,m}$ $<$ 0.25), and mini mergers ($M_{\rm \star,c}/M_{\rm \star,m}$ $<$ 0.05).


\section[]{S0 formation pathways}
\label{sec: S0 formation pathways}

The relationship between gas content and galactic morphology has been well studied, revealing that ETGs typically have lower gas-to-stellar mass fractions \citep[][]{robert1994morphSF, calette2018Hmassfraction}. While the origin of spiral arms remains a subject of active debate, gas is expected to play an important role in building spiral structures through the process of swing amplification \citep[][]{sellwood1984spiral, donghia2013perturber}. In the absence of a gas supply, a dynamically cold stellar disk cannot be maintained, leading instead to a heated and thickened disk via orbital diffusion, which prevents the development of spiral features \citep[][]{toomre1964toomreQ, fouvry2017diskheating}. On this account, understanding the history of gas content in galaxies is crucial in studying galactic morphology.

One possible pathway leading to poor gas content in galaxies is the cancellation of gas angular momentum.  When the angular momentum of gas decreases, enhanced inflows drive gas from the outer regions toward the center, where it is rapidly exhausted through star formation due to high gas density \citep[][]{lovelace1996counterrot, thakar1998counterrot, pizzella2004counter, chen2016bulgegrowth, beom2022counterrot, zhou2023misalinged}. Angular momentum loss in gas can occur via various channels: tidal interactions during galaxy-galaxy encounters \citep[][]{barnes1991gasAMloss, moore1996encounter, lambas2003galaxyinteraction}, tidal torques exerted from clumps formed via gravitational instability \citep[][]{noguchi1999clump, elmegreen2012gasAMloss, zolotov2015gasAMloss}, and hydrodynamic collisions between pre-existing gas and accreted counter-rotating gas \citep[][]{danovich2015gasAMloss, dyda2015counterrotatingaccretion}. Notably, central gas inflow driven by angular momentum loss is a critical process in the formation pathways of S0 galaxies, which we explore in the following sections.

In this section, we examine the evolutionary histories of two quenched S0 galaxies to explore how their morphology develops alongside the changes in physical properties, particularly those related to gas content. The range of redshift we investigate spans from $z=4$ to $z=0.17$.

\subsection{S0\,19: Merger}
\label{sec: S0 formation pathways - S0 19: Merger}

\subsubsection{Gas angular momentum history}
\label{sec: S0 formation pathways -  S0 19: Mergers - Gas angular momentum history}

\begin{figure}
\centering
\includegraphics[width=0.5\textwidth]{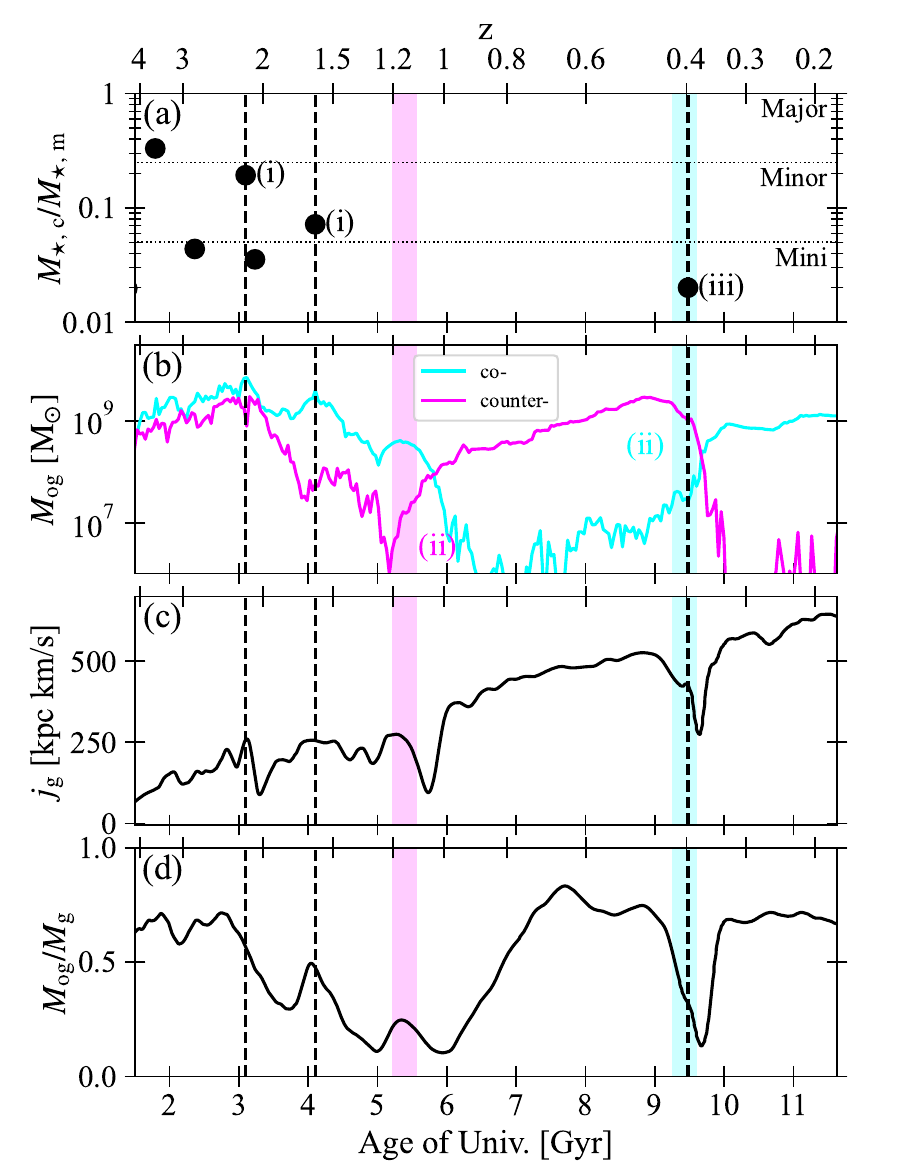}
\caption{Merger history and evolution of gas properties of S0\,19.
The black points in (a) indicate the occurrence time and mass ratio of merger events.
The two horizontal dotted lines correspond to mass ratios of 1:4 and 1:20.
The three mergers mentioned in the text are indicated by (i) and (iii) with vertical dashed lines across all panels.
(b) shows the evolution of the mass of co-rotating (cyan) and counter-rotating gas (magenta) in the outer region ($\re < R < 2\re$).
The two gas accretion events explained in the text are indicated by (ii) with shaded regions in all panels.
The widths of the shaded regions correspond to the times during which the masses of the co- or counter-rotating gas straddle between 1/100 and 1/10 of $M_{\rm og}$, indicating an increase in gas content.
(c) and (d) show the evolution of $j_{\rm g}$ and outer gas mass fraction, respectively.
The lines in (c) and (d) are smoothed by an Epanechnikov kernel with a time interval of 20 simulation snapshots ($\sim300$\,Myr).
}
\label{fig04_S019_gas_j}
\end{figure}

Figure~\ref{fig04_S019_gas_j}(a) and (b) respectively show the merger history and the mass evolution of co-rotating and counter-rotating gas of S0\,19. To define co-rotating and counter-rotating gas, we utilize the total stellar rotation axis vector, the summation of angular momentum vectors of stellar particles within 2\re, measured with respect to the galactic center. A gas cell is classified as co-rotating if the angular difference between its rotation axis vector and the stellar rotation axis vector ($\theta$) is less than 90 degrees and counter-rotating if $\theta$ is greater than 90 degrees. Both gas components are defined according to the stellar rotation in each snapshot. To calculate their mass, we use gas cells in the ``outer region'', $\re < R < 2\re$, where $R$ is the radial distance from the galactic center. 
The exclusion of the inner region ($< \re$) in the analysis helps minimize the effects of hydrodynamic mixing, which can complicate the tracing of original gas motions.

\begin{figure*}
\centering
\includegraphics[width=0.9\textwidth]{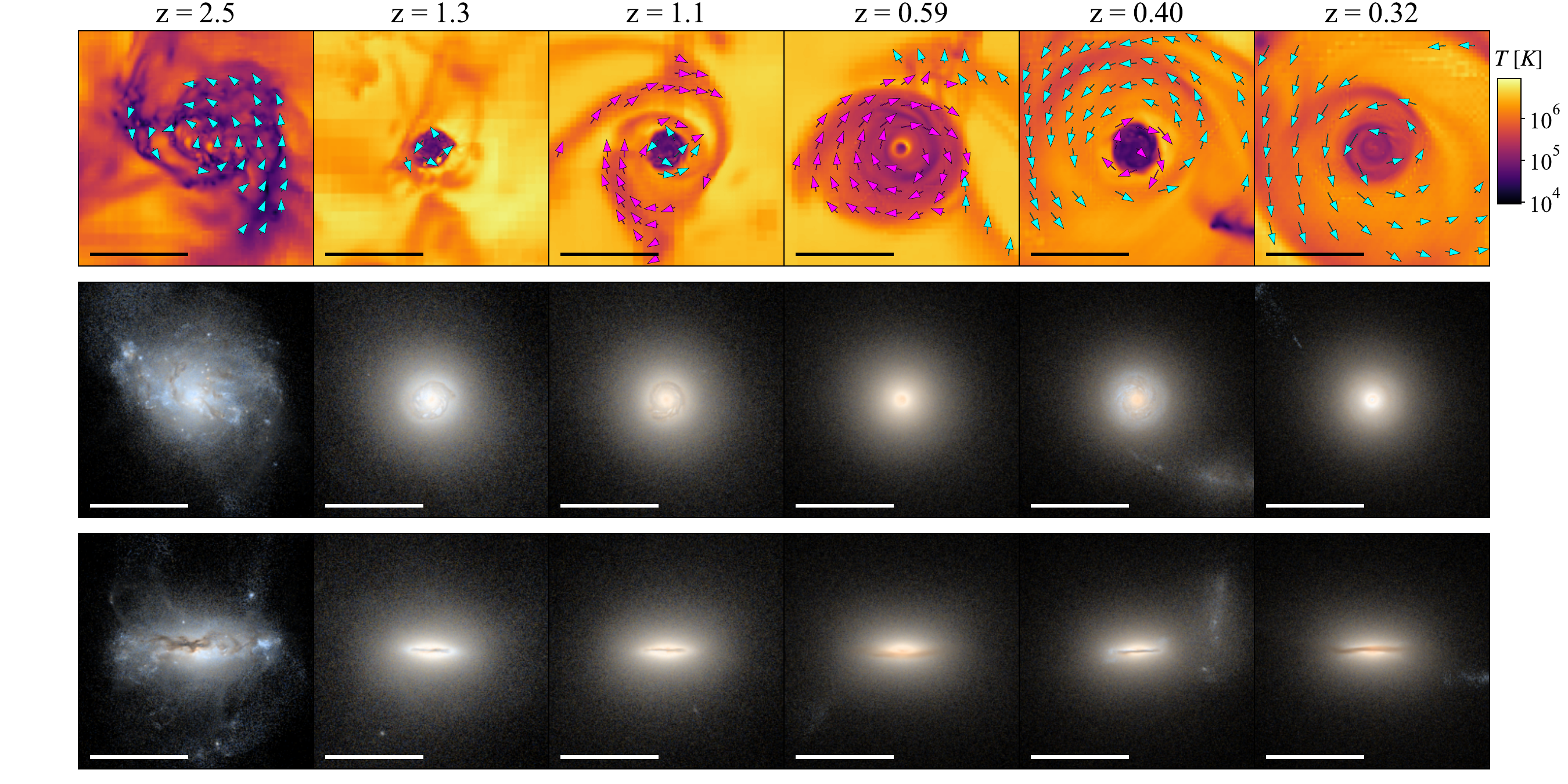}
\caption{Gas temperature maps and mock images of S0\,19 at different redshifts.
The images in the same column correspond to the same redshift indicated above.
The gas temperature maps are shown in the top row, with colors representing the projected density-weighted temperature.
The mock images viewed face-on and edge-on are shown in the middle and bottom rows, respectively.
The black and white bars in the panels indicate 10\,kpc.
The cyan and magenta arrows in the gas maps represent the projected velocity of co-rotating and counter-rotating gas, respectively.
After two minor mergers, the gas disk becomes highly compact, and the galaxy quickly turns into an S0, exhibiting a smooth morphology ($z = 2.5$ and 1.3).
The alternating co-rotating and counter-rotating gas accretions interact with the pre-existing gas disk, impeding the development of the gas disk ($z = 1.1$, 0.59, and 0.40).
A ``mini'' merger leads to a rapid shrinkage of the gas disk ($z = 0.40$ and 0.32).
 }
\label{fig05_S019_mockimgs}
\end{figure*}

Our major concern is whether individual gas cells lose angular momentum and migrate toward the center. To see this, we use the gas cells within 2\re~and compute the sum of their angular momentum divided by the total gas mass ($j_{\rm g}$) at each epoch in the following way:
\begin{equation}
    j_{\rm g} = \frac{\sum_{i}|\vec{J_{i}}|}{\sum_{i}{M_{i}}} \, ,
\label{eq2_j_g}
\end{equation}
where $\vec{J_{i}}$ and ${M_{i}}$ are the angular momentum and mass of the $i^{\rm th}$ gas cell, respectively. Unlike the conventional definition of specific angular momentum (${\sum_{i}\vec{J_{i}}}/{\sum_{i}{M_{i}}}$), which involves vector summation, $j_{\rm g}$ quantifies the magnitude of motion of gas cells regardless of rotation direction. For S0\,19, Figure~\ref{fig04_S019_gas_j}(c) shows the evolution of $j_{\rm g}$, while Figure~\ref{fig04_S019_gas_j}(d) shows the evolution of the fraction of outer gas mass, $M_{\rm og}/M_{\rm g}$, where $M_{\rm og}$ is the total mass of gas cells in the outer region and $M_{\rm g}$ is that of gas cells within 2\re. A simultaneous decrease in $j_{\rm g}$ and $M_{\rm og}/M_{\rm g}$, which is visible at several epochs, suggests that the cancellation of gas angular momentum drives inward gas flow, enhancing the spatial concentration of the gas (``gas concentration'').

\textbf{(i) Minor mergers}

In Figure~\ref{fig04_S019_gas_j}, the early gas concentration is seen during the two minor mergers: one with $M_{\rm \star,c}/M_{\rm \star,m}\simeq 0.19$ at $z\sim2.3$ and another with $M_{\rm \star,c}/M_{\rm \star,m}\simeq 0.07$ at $z\sim1.6$. The outer gas mass ratio drops to around 0.1 as a result of two subsequent minor mergers, indicating that most of the gas is concentrated near the center. Figure~\ref{fig05_S019_mockimgs} shows gas temperature maps and galaxy mock images of S0\,19 at different redshifts. At $z=2.5$ and 1.3, the gas maps show that the gas distribution becomes more centralized due to the merger events, with a significant reduction in gas at the outskirts. Accordingly, the mock images exhibit the transition from a clumpy and disturbed appearance to a smoother, S0-like morphology. Meanwhile, the stellar spin (in $V/\sigma_{\star}$) remains strong even after the mergers (see Appendix~\ref{sec: Appendix - Stellar kinematics}), suggesting that the galaxy retains both kinematic and visual disk-like structures.

\textbf{(ii) Gas accretion in the opposite direction}

In Figure~\ref{fig04_S019_gas_j}(b), the mass of counter-rotating gas increases from $z\sim1.2$, while the co-rotating gas mass decreases, leading to an inversion of the primary and secondary gas components. At $z\sim0.4$, this trend reverses, with the co-rotating gas mass increasing again. These shifts are due to gas accretion in the opposite direction to the rotation of the primary gas component, as shown in gas maps from $z=1.1$ to 0.4 in Figure~\ref{fig05_S019_mockimgs}. At $z=1.1$, the counter-rotating gas (magenta arrow) accumulates near the periphery of the co-rotating gas (cyan arrow) disk as it spins around the galaxy. The mixing of the two gas components results in angular momentum cancellation and gas inflow, creating a void boundary between two gas disks ($z=0.59$) \citep[e.g.,][]{dyda2015counterrotatingaccretion}. As a result, the pre-existing co-rotating gas disk shrinks and forms stars. Subsequently, additional co-rotating gas is accreted, playing a role similar to that of the previous counter-rotating gas ($z=0.4$). This alternating gas accretion hinders the mass and size growth of the gaseous disk. We visually find that the accreted gas in this case is coming from the stripped gas of passing satellites or from free-floating gas rather than from gaseous filaments.

\textbf{(iii) Mini merger}

In Figure~\ref{fig05_S019_mockimgs} at $z=0.4$, the merging companion is visible at the right, and the galaxy mock image exhibits a bluish color due to the star formation of the counter-rotating gas disk described in the previous section. Following the mini merger ($M_{\rm \star,c}/M_{\rm \star,m}\simeq 0.02$) with the companion at $z\sim0.4$, the counter-rotating gas disk rapidly shrinks to disappear, resulting in the final S0 morphology ($z=0.32$). The dramatic gas concentration during this period is also illustrated in Figure~\ref{fig04_S019_gas_j}.

\subsubsection{Quenching processes}
\label{sec: S0 formation pathways -  S0 19: Mergers - Quenching processes}

\begin{figure}
\centering
\includegraphics[width=0.5\textwidth]{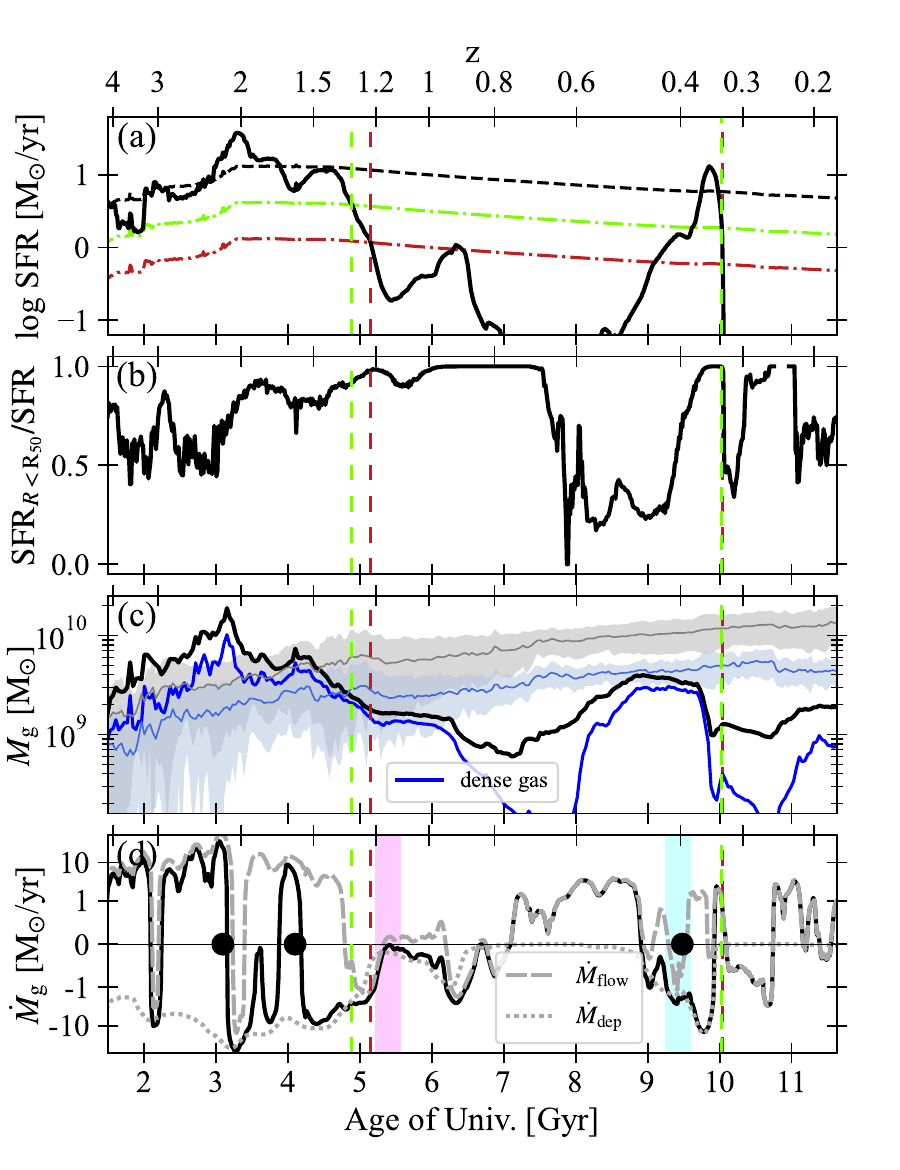}
\caption{Temporal evolution of the SFR and gas content of S0\,19.
(a) shows SFR as a function of time. The black dashed line represents the SFMS according to Equation~(\ref{eq1_SFMS}) derived for the stellar mass of S0\,19 at each redshift.
The green and red dot-dashed lines indicate 0.5 and 1.0 dex below the SFMS, respectively.
The green and red vertical lines in all panels indicate $t_{\rm BQ}$ and $t_{\rm Q}$, respectively.
(b) displays the evolution of the fraction of SFR within \re.
(c) compares the gas content history of S0\,19 with its late-type counterparts. The $M_{\rm g}$ of S0\,19 is represented as a black solid line.
The mean and $1\,\sigma$ range of $M_{\rm g}$ of the late-type counterparts are shown as the gray solid line and shaded area, respectively.
The same information for dense gas is presented in blue colors.
(d) shows the history of $\dot M_{\rm g}$ (black solid line), $\dot M_{\rm dep}$ (gray dotted line), and $\dot M_{\rm flow}$ (gray dashed line).
The lines are smoothed by an Epanechnikov kernel with a size of 20 simulation snapshots.
The points and shaded regions represent the same events highlighted in Figure~\ref{fig04_S019_gas_j} to show their impacts on gas content.
At $z\sim1.2$ and $z\sim0.33$, star formation is quenched after significant gas consumption due to mergers within $\tau_{\rm Q}$ of 270\,Myr and 15\,Myr, respectively.
}
\label{fig06_S019_quenching}
\end{figure}

Now, we investigate how S0\,19 is quenched through the events discussed earlier. Figure~\ref{fig06_S019_quenching} presents the temporal evolution of the SFR and gas content of S0\,19. Figure~\ref{fig06_S019_quenching}(a) shows the SFR as a function of time alongside the SFMS following Equation~(\ref{eq1_SFMS}) corresponding to its \mstellar~at each redshift. To assess the impact of gas concentration on the increase in central SFR, we present the fraction of SFR measured inside \re~(${\rm SFR}_{R<\re}$) relative to the total SFR as a function of time in Figure~\ref{fig06_S019_quenching}(b). Figure~\ref{fig06_S019_quenching}(c) shows the evolution of the total mass of all gas ($M_{\rm g}$) and dense gas inside 2\re. Here, dense gas cells are those with $n_{\rm H} > 10\,\rm cm^{-3}$, where star formation is likely according to the star formation efficiency of the cells. For comparison, we use seven late-type galaxy counterparts whose \mstellar~differ by less than 0.25\,dex (in log scale) from that of S0\,19 at $z=0.17$. 

Figure~\ref{fig06_S019_quenching}(d) shows the rate of $M_{\rm g}$ change over time ($\dot M_{\rm g}$)$\text{---}$the difference in $M_{\rm g}$ of two consecutive snapshots divided by the time interval. Additionally, we utilize the gas mass depletion rate ($\dot M_{\rm dep}$) and net gas mass flow ($\dot M_{\rm flow}$) to trace the causes of changes in gas mass content. To compute $\dot M_{\rm dep}$, we calculate the amount of gas used up for star formation at each epoch by identifying stellar particles born within 2\re. As stellar particles return $\sim31\%$ of their initial mass in the form of gas, we consider that the remnant stellar mass (i.e., $\sim69\%$ of the initial mass of stellar particles) is depleted inside 2\re. Since we can directly calculate $\dot M_{\rm g}$ and $\dot M_{\rm dep}$, we obtain $\dot M_{\rm flow}$ using the following relation:
\begin{equation}
    \dot M_{\rm g} = \dot M_{\rm flow} + \dot M_{\rm dep}.
\label{eq3_dotMg}
\end{equation}
Thus, $\dot M_{\rm flow}$ indicates the net gas mass flow at the spherical boundary of 2\re. We ignore gas consumed by BH sink particles since it is negligible ($< \rm 10^{-3} \, \msol/yr$).

In Figure~\ref{fig06_S019_quenching}(d), the negative $\dot M_{\rm g}$ occurs immediately following each minor merger (black points), primarily due to significant gas depletion by star formation. Since the gas mass content drops quickly, as shown in Figure~\ref{fig06_S019_quenching}(c), the SFR becomes quenched subsequently over $\tau_{\rm Q}$ of 270\,Myr. Notably, Figure~\ref{fig06_S019_quenching}(b) indicates that this substantial gas consumption during mergers is spatially concentrated, highlighting the effects of central gas inflow. Next, counter-rotating gas accretion indicated by the magenta shade in Figure~\ref{fig06_S019_quenching}(d) induces gas concentration, which leads to further star formation and gas depletion in turn at $z\sim1$. As a result, the galaxy has a poor gas mass content.

In Figure~\ref{fig06_S019_quenching}, co-rotating gas accretion steadily increases the SFR prior to the mini merger at $z\sim0.4$. The final notable gas depletion is observed when the gas disk quickly contracts due to the mini merger, leading to a second instance of galaxy quenching. This quenching process is extremely rapid, occurring within a single snapshot interval ($\sim15\,$Myr). Thus, we argue that gas concentration caused by mergers contributes to star formation quenching through substantial gas consumption, predominantly near the center.

\subsection{S0\,26: Counter-rotating gas accretion}
\label{sec: S0 formation pathways - S0 26: Counter-rotating gas accretion}

The effect of retrograde gas acquisition (i.e., gas-gas interaction triggers central gas inflow) has been highlighted in the previous section. Here, we investigate how counter-rotating gas accretion can function as a primary pathway for S0 formation by examining the evolutionary history of S0\,26.

\subsubsection{Gas angular momentum history}
\label{sec: S0 formation pathways - S0 26: Counter-rotating gas accretion - Gas angular momentum history}

\begin{figure}
\centering
\includegraphics[width=0.5\textwidth]{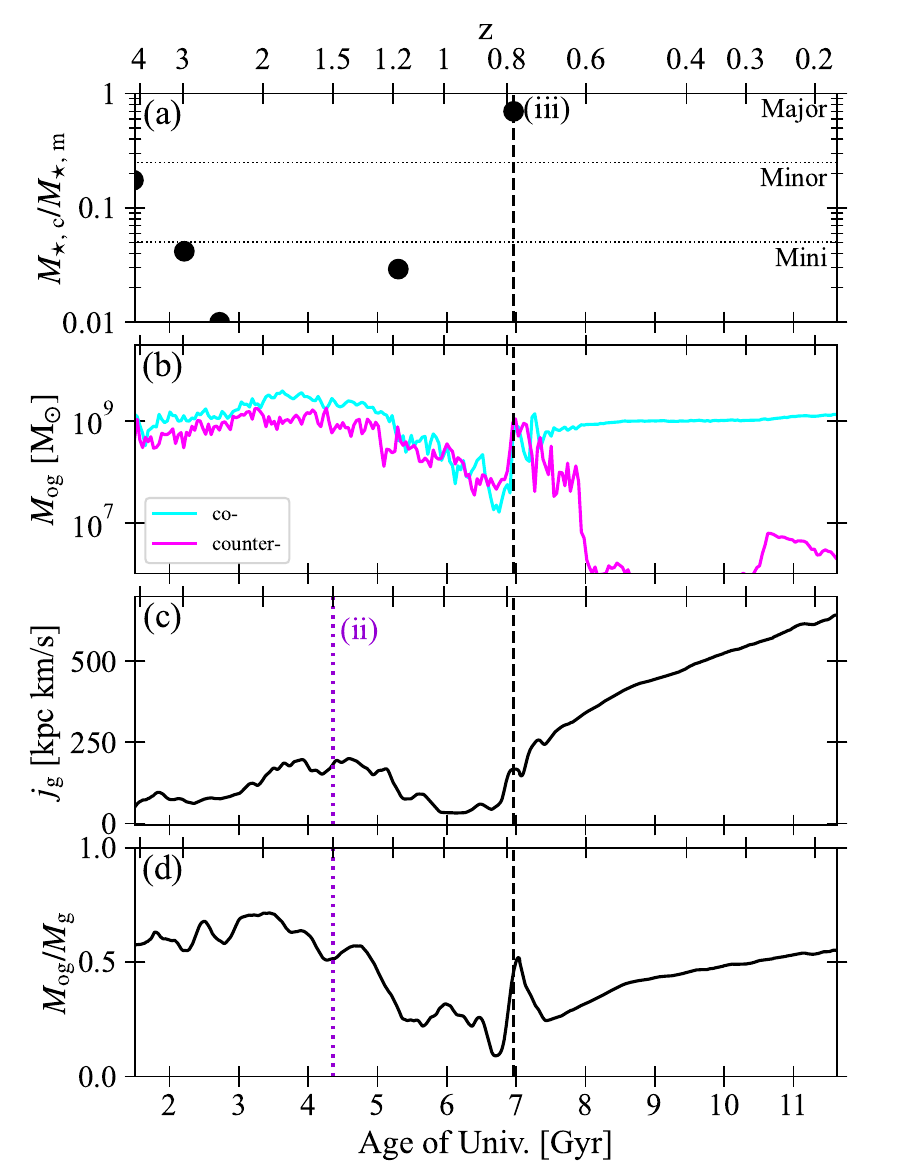}
\caption{
Merger history and evolution of the gas properties of S0\,26 in the same format as Figure~\ref{fig04_S019_gas_j}.
The beginning of the gas concentration mentioned in the text is indicated by the vertical line marked with (ii).
Similarly, the major merger at $z\sim0.8$ is marked by (iii).
}
\label{fig07_S026_gas_j}
\end{figure}

\begin{figure}
\centering
\includegraphics[width=0.5\textwidth]{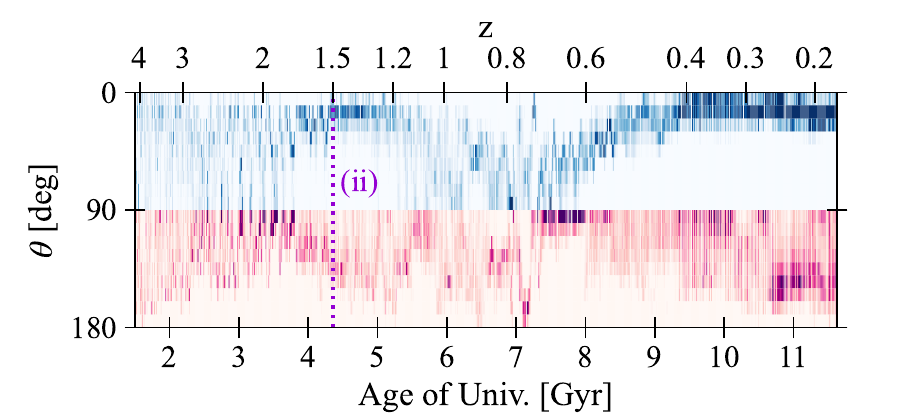}
\caption{The distribution of $\theta$ for gas cells measured inside 2\re~of S0\,26, as a function of time.
The blue and pink colors represent the co- and counter-rotating gas cells, respectively.
The colors of each gas component are weighted by the magnitude of the angular momentum of gas cells at each redshift; thus, darker colors indicate that more gas cells are rotating in the direction of $\theta$ with greater angular momentum.
The onset of the gas concentration is indicated in the same manner as Figure~\ref{fig07_S026_gas_j}.
 }
\label{fig08_S026_gas_theta}
\end{figure}

\begin{figure*}
\centering
\includegraphics[width=0.9\textwidth]{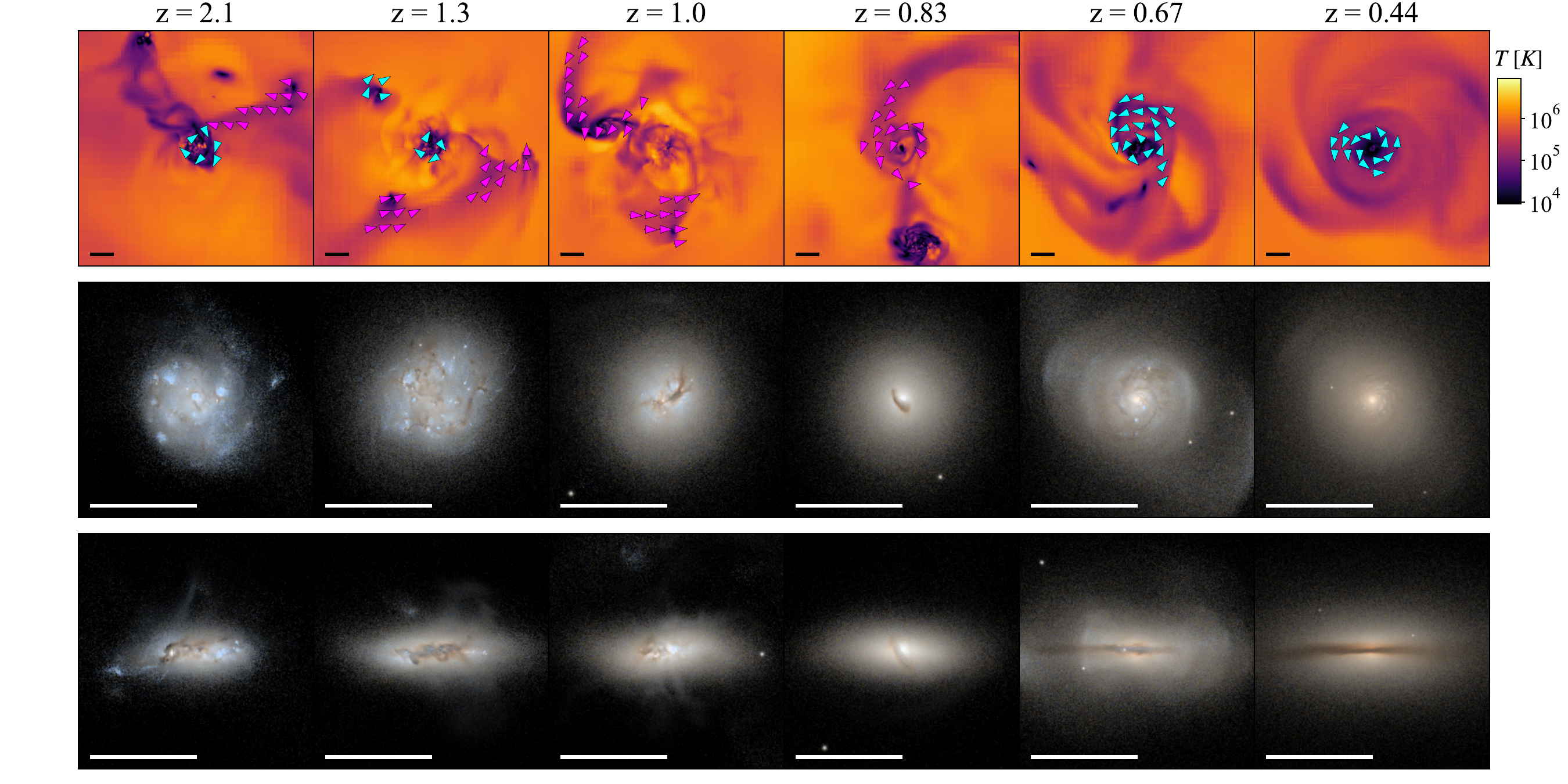}
\caption{Gas temperature maps and mock images of S0\,26 in the same format as Figure~\ref{fig05_S019_mockimgs}.
The galaxy acquires counter-rotating gas via smooth accretion ($z=2.1$) and from small satellites ($z=1.3$).
Tidal interaction induces the two-armed spirals in the companion galaxy, and the stripped gas is accreted as counter-rotating gas due to the retrograde orbit of the merger ($z=1.0$).
An S0 morphology appears due to the lack of gas in the outskirts ($z=0.83$).
After experiencing the major merger, shell-like structures are visible, and the reversed stellar rotation results in the switch between two gas components as co- and counter-rotating ($z=0.67$).
The merger does not destroy the visual stellar disk, although it comprises two oppositely rotating stellar components ($z=0.44$).
}
\label{fig09_S026_mockimgs}
\end{figure*}

\textbf{(i) Early existence of counter-rotating gas}

Figure~\ref{fig07_S026_gas_j} shows the merger history and evolution of gas properties associated with gas concentration for S0\,26. In Figure~\ref{fig07_S026_gas_j}(b), a comparable amount of counter-rotating gas to co-rotating gas continues since $z\sim4$. 
Unlike the case of the previous galaxy (S0\,19), however, we do not see gas concentration until $z\sim1.5$ (Figure~\ref{fig07_S026_gas_j}(c) and (d)).

Figure~\ref{fig08_S026_gas_theta} shows the distribution of $\theta$ for co-rotating (blue) and counter-rotating gas (pink) as a function of time. As a reminder, $\theta$ is the angle between the rotation axes of gas and stars. For both gas components, darker colors indicate that there are more gas cells with high angular momentum rotating at a given $\theta$. In the early epoch ($z\gtrsim2$), both gas components show non-aligned rotation, reducing the chance of effective collisions between them.

\textbf{(ii) Beginning of gas concentration}

Figure~\ref{fig08_S026_gas_theta} shows that, while the counter-rotating gas remains misaligned with the stellar disk, the co-rotating gas becomes quickly aligned with the stellar disk ($\theta$ approaching 0) after $z\sim1.5$. This enhances the hydrodynamic collision between the two gas components, leading to angular momentum cancellation. As a result, a notable gas concentration develops at $z\sim1.5$ as shown in Figure~\ref{fig07_S026_gas_j}(c) and (d).

The gas maps in Figure~\ref{fig09_S026_mockimgs} display that the galaxy gains gas in retrograde orbits through the smooth accretion of cold gas ($z=2.1$) and the accretion of small satellite galaxies ($z=1.3$).

\begin{figure}
\centering
\includegraphics[width=0.5\textwidth]{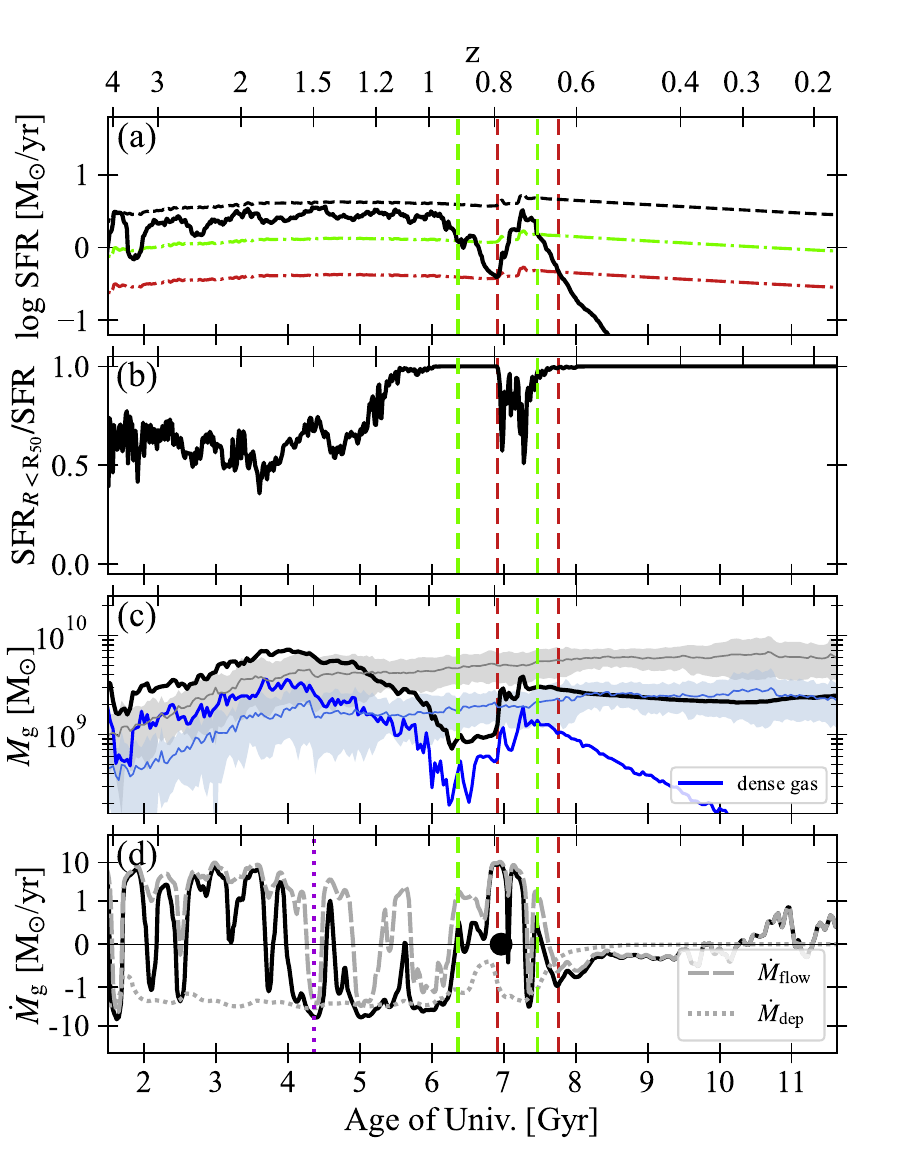}
\caption{Temporal evolution of the SFR and gas content of S0\,26 in the same format as Figure~\ref{fig06_S019_quenching}. 
Fifteen late-type counterparts are used for comparison in (c).
The black point in (d) represents the time when the major merger occurs.
The first star formation quenching at $z\sim0.8$ is followed by the concentration of gas via hydrodynamic interactions between co- and counter-rotating gas ($\tau_{\rm Q}$ = 550\,Myr).
Despite the rejuvenation through the major merger, the galaxy is quenched again ($\tau_{\rm Q}$ = 280\,Myr).}
\label{fig10_S026_quenching}
\end{figure}

\textbf{(iii) Major merger}

The gas map in Figure~\ref{fig09_S026_mockimgs} at $z=1.0$ shows that the galaxy undergoes a tidal interaction with a major merging companion at the pericenter. Gravitational torques exerted by this encounter drive prominent two-armed spirals in the companion galaxy \citep[e.g.,][]{toomre1972spiral, dobbs2011spiral}. Moreover, due to the retrograde orbit of the merger, tidally stripped gas is provided to the galaxy as additional counter-rotating gas during this process. Figure~\ref{fig07_S026_gas_j} demonstrates the lowest $j_{\rm g}$ and outer gas mass fraction at $1 \lesssim z \lesssim 0.8$ as a result of the continuous gas-gas interaction since $z\sim1.5$ and gravitational torques prior to the merger. Consequently, the galaxy’s morphology evolves to a smoother form, as shown in Figure~\ref{fig09_S026_mockimgs} at $z=1.0$ and 0.83.

Despite the major merger at $z\sim0.8$, the stellar disk remains visible at $z=0.44$ in Figure~\ref{fig09_S026_mockimgs} because the slow coplanar merger proceeds in a non-destructive manner \citep[e.g.,][]{zeng2021majormerger}. However, the total stellar rotation axis reverses as the combined angular momentum of the newly incorporated stars and pre-existing counter-rotating stars exceeds that of co-rotating stars. As a result, the gas that once counter-rotated is recognized as the co-rotating component (Figure~\ref{fig07_S026_gas_j}(b)). Two stellar disks rotating in opposite directions on the plane counteract each other's rotational velocities, making the galaxy a kinematically pressure-supported system (see Appendix~\ref{sec: Appendix - Stellar kinematics}).

\subsubsection{Quenching processes}
\label{sec: S0 formation pathways - S0 26: Counter-rotating gas accretion - Quenching processes}

The effects of gas concentration, beginning at $z\sim1.5$, are evident in Figure~\ref{fig10_S026_quenching}. As the gas distribution becomes more concentrated, star formation also becomes more spatially focused (Figure~\ref{fig10_S026_quenching}(b)). Despite the overall decline in gas mass content, the SFR remains steady because the high gas density at the center facilitates efficient star formation (Figure~\ref{fig10_S026_quenching}(a) and (c)). Consequently, the significant reduction in gas mass due to the gas consumption leads to the quenching of star formation at $z\sim0.8$ ($\tau_{\rm Q}$ = 550\,Myr). It is important to note that the tidal interaction with a companion at $z\sim1$ may also contribute to the quenching process by triggering further central gas inflow, making it challenging to attribute the quenching solely to counter-rotating gas accretion.

The major merger temporarily rejuvenates the galaxy; however, the increased star formation is short-lived. The galaxy is quenched again over $\tau_{\rm Q}$ of 280\,Myr. Subsequently, the dense gas gradually depletes, resulting in the absence of star formation. The lack of gas inflow following the major merger (Figure~\ref{fig10_S026_quenching}(d)) allows the galaxy to maintain its S0 morphology. In this case, the absence of subsequent gas supply does not serve as the main driver for S0 morphology but prevents it from re-establishing late-type morphology.

%

\section[]{Discussion}
\label{sec: Discussion}

\subsection{Gas angular momentum cancellation mechanisms}
\label{sec: Discussion - Gas angular momentum cancellation mechanisms}

By investigating the formation history of the two NH S0 galaxies in the previous section, we find that central gas migration due to a decrease in gas angular momentum is a key factor in the development of S0 morphology. The evolution of S0\,19 and S0\,26 provides evidence for mergers and counter-rotating gas accretion as distinct pathways to S0 formation. The physical mechanisms by which gas angular momentum is cancelled vary depending on the scenarios.

\subsubsection{Mergers}
\label{sec: Discussion - Gas angular momentum cancellation mechanisms - Mergers}

During mergers between galaxies with similar mass, an asymmetric structure is produced due to a strong tidal field; such asymmetry can generate a gravitational torque, which deprives the angular momentum of gas \citep[e.g.,][]{hernquist1989majormerger, barnes1991gasAMloss, mihos1996majormerger, di2007majormerger}. Minor mergers can generate a hydrodynamic torque induced by a ram-pressure shock, resulting in the angular momentum deprivation \citep[][]{capelo2017rampressureshock}. In this case, the merger geometry may influence this process since coplanar encounters facilitate hydrodynamic interaction between the two gas disks, and retrograde encounters strengthen ram pressure with higher relative velocity \citep[][]{capelo2017rampressureshock, blumenthal2018mergergeometry}. Through these mechanisms, nuclear gaseous inflows develop. The consequences are well demonstrated in S0\,19 even during a mini merger, implying that a merger of such a small mass ratio may drive noticeable effects.

\subsubsection{Counter-rotating gas accretion}
\label{sec: Discussion - Gas angular momentum cancellation mechanisms - Counter-rotating gas accretion}

Within a gas-gas counter-rotating galaxy, the two different gas components with opposite spin directions collide; this hydrodynamic interaction leads to gas angular momentum cancellation. As shown in Section~\ref{sec: S0 formation pathways - S0 26: Counter-rotating gas accretion - Gas angular momentum history}, this process can be more effective for coplanar rotation since it maximizes interactions along the orbits. If a substantial amount of gas falls toward the center, the insufficient amount of outer gas may suppress the growth of spiral arms. The resulting high central gas density triggers efficient star formation, possibly contributing to the bulge growth and a younger stellar population in the center compared to the outskirts \citep[][]{chen2016bulgegrowth, zhou2023misalinged}.

\subsubsection{Environmental dependence}
\label{sec: Discussion - Gas angular momentum cancellation mechanisms - Environmental dependence}

We now discuss how mergers and counter-rotating gas accretion may be appropriate mechanisms for the formation of S0 galaxies, particularly in low-density environments.

Counter-intuitively, the suitable environments for frequent galaxy-galaxy mergers are small groups rather than dense clusters, as high relative velocities of cluster galaxies lower the chance of collisions \citep[][]{ostriker1980clustermerger}. Observational studies support this by showing a decline in merger fractions in higher-density environments \citep[][]{delahaye2017mergerfrac, pearson2024mergerfrac}, although preprocessing of galaxies (i.e., interactions in previous haloes) can blur this trend \citep[e.g.,][]{sheen2012prepross, oh2018prepross}. \cite{just2010S0frac} proposed that mergers can be an important S0 formation channel in environments with lower velocity-dispersion, at least since $z \sim 0.8$. \cite{wilman2009fieldS0} revealed that groups are more suitable environments for galaxy mergers than fields where internal secular processes are more likely to occur.

Kinematically misaligned galaxies are more frequently observed in field environments \citep[][]{davis2011denseenv, jin2016counterrotation}. This implies that external gas accretion onto galaxies, which may occur isotropically, is less obstructed in these low-density settings, whereas it can be prevented in dense environments via stripping of extended gas haloes or shock heating of cold gas flow \citep[][]{larson1980strangulation, dekel2006shutdown, bahe2012deneseenv}. In that sense, the counter-rotating gas accretion scenario is likely to be preferred by S0s in low-density environments. Various counter-rotating gas channels are suspected, such as mergers and smooth accretion from gaseous filaments \citep[][]{taylor2018originofmisgas, khim2021misaglinment}.

\subsection{Feedback enhancement}
\label{sec: Discussion - Feedback enhancement}

\begin{figure}
\centering
\includegraphics[width=0.5\textwidth]{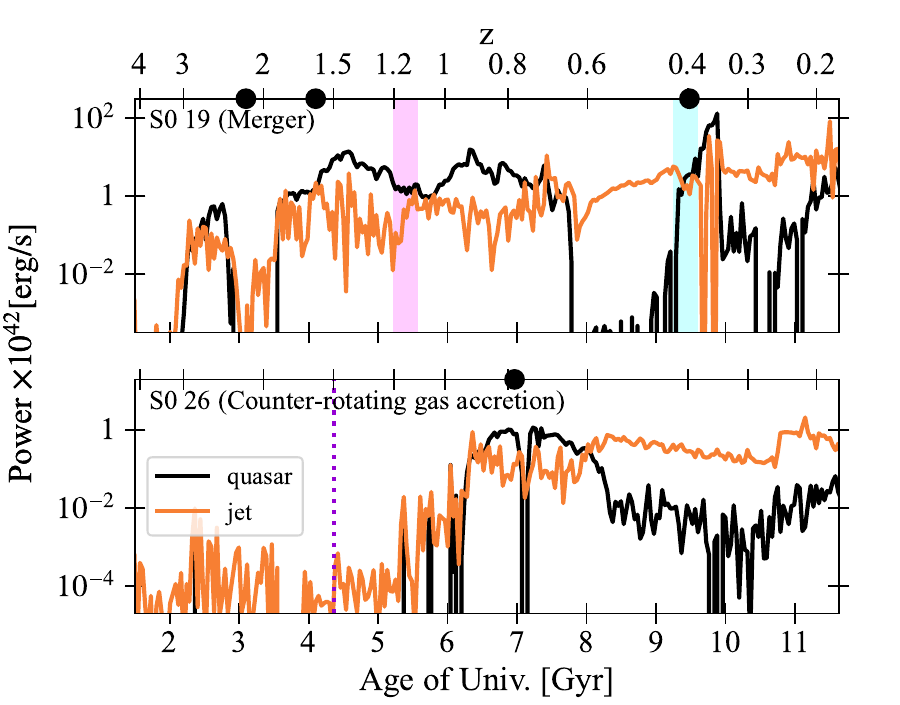}
\caption{Evolution of the power of AGN feedback for S0\,19 and S0\,26.
The power of quasar mode (black) and jet mode (orange) feedback is measured from BHs within \re.
The black points and shaded areas indicate the merger and gas accretion events mentioned in the previous section.
S0\,19 and S0\,26 each possess the most massive BH with a mass of $\rm 10^{7.2}\, M_{\odot}$ and $\rm 10^{5.7}\, M_{\odot}$, respectively, at $z=0.17$.
}
\label{fig11_S019_S026_AGN}
\end{figure}

The impact of mergers and misaligned gas accretion on BH fueling via central gas inflow has been well documented by numerical simulations and observations \citep[e.g.,][]{springel2005mergerquench, capelo2015mergerAGN, hong2015mergerAGN, beom2022counterrot, raimundo2023misaligendAGN}. To determine whether this also occurs during the evolution of S0\,19 and S0\,26, we present Figure~\ref{fig11_S019_S026_AGN} showing the power of AGN feedback measured from the BHs of the two S0 galaxies as a function of time. For the BHs within \re, we calculate the power of AGN feedback by considering the total energy released between consecutive snapshots and the time interval. Although a BH sink particle can produce feedback energy of only one mode (kinetic energy from jet mode or thermal energy from quasar mode), the feedback of two modes can be seen at the same time if there are multiple BHs inside \re, which is very likely in merging conditions.

For S0\,19, the power of AGN feedback diminishes during the first minor merger. We visually confirm that the BHs move away from the galactic center at that time due to the fluctuation of the gravitational potential. Consequently, they accrete less gas but eventually settle at the galactic center. AGN feedback is enhanced following the second minor merger. The counter-rotating gas accretion also plays a role in the enhancement, albeit with a time delay. Following the accretion, as the mixing of two gas components contributes to the shrinkage of the pre-existing gas disk, more gas is driven to the center. The highest power of AGN feedback is visible after the mini merger since the counter-rotating gas disk rapidly shrinks and fuels BHs. This is well illustrated in a separate NH paper, which found that the spin direction of the most massive BH in S0\,19 is reversed due to the substantial accretion of counter-rotating gas onto the BH following the mini merger \citep[][BH-1049 therein]{peirani2024BHspin}.

For S0\,26, the power of AGN feedback increases as the gas concentration begins at $z\sim1.5$. The feedback strength continues to grow via the tidal interaction at $z\sim1$ and the coalescence at $z\sim0.8$.

\begin{figure}
\centering
\includegraphics[width=0.5\textwidth]{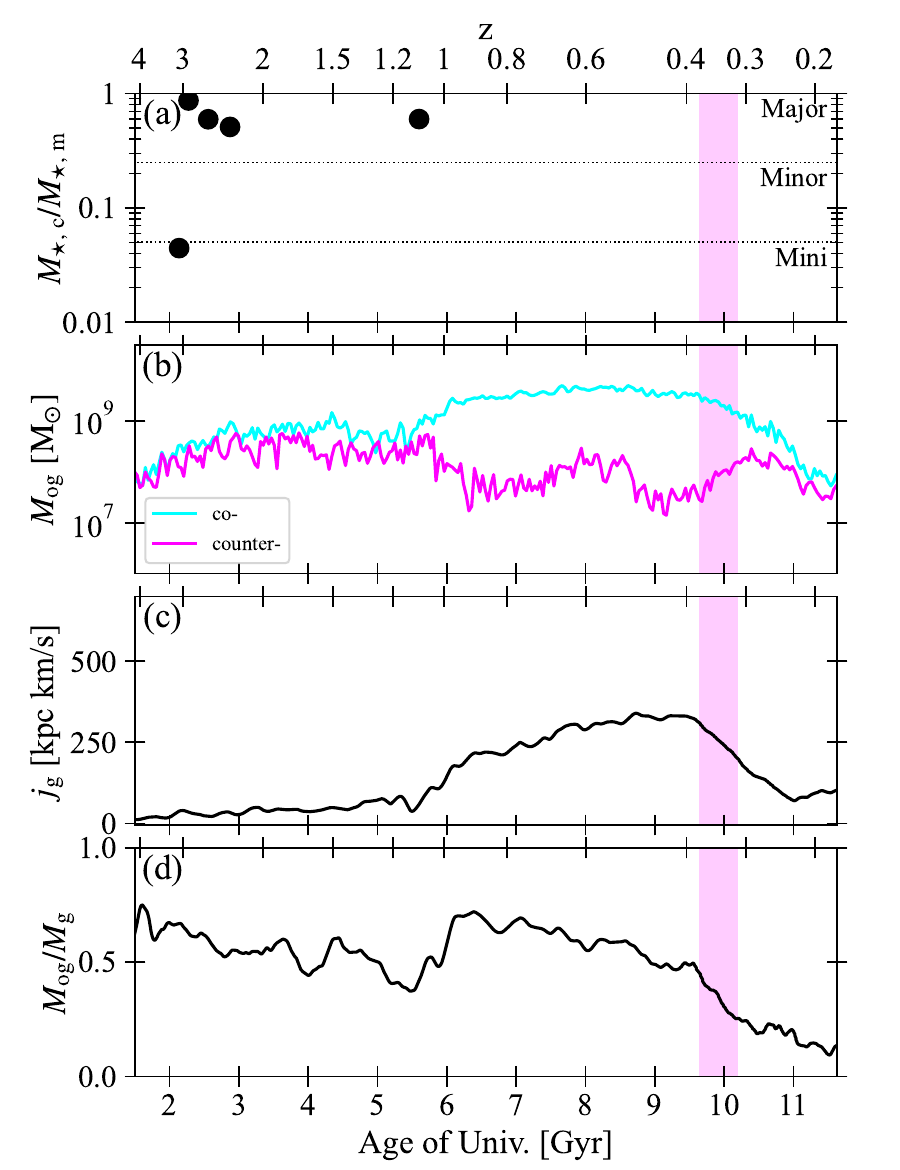}
\caption{Merger history and evolution of the gas properties of S0\,21 in the same format as Figure~\ref{fig04_S019_gas_j}. 
A notable gas concentration occurs as counter-rotating gas mass increases in the outer region (near the vertical pink band), implying the gas concentration process is in place.}
\label{fig12_S021_gas_j}
\end{figure}

\begin{figure}
\centering
\includegraphics[width=0.5\textwidth]{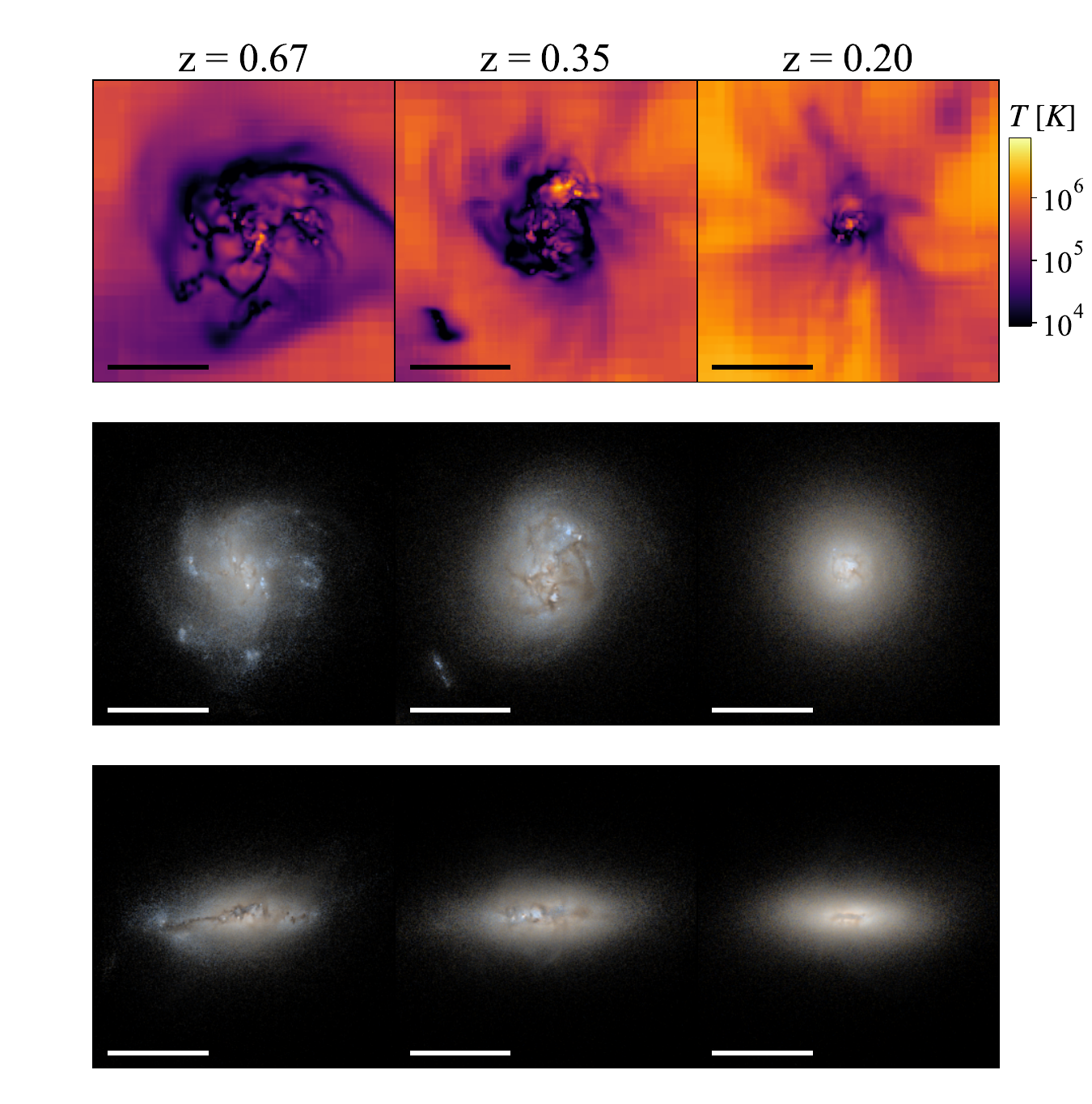}
\caption{Gas temperature maps and mock images of S0\,21 in the same format as Figure~\ref{fig05_S019_mockimgs}, but for three snapshots.
The gas distribution becomes more concentrated due to the interaction between the pre-existing and counter-rotating gas.
The morphology transforms in line with the gas distribution, exhibiting more S0-like morphology.}
\label{fig13_S021_mockimgs}
\end{figure}

\subsection{Quenching timescale}
\label{sec: Discussion - Quenching timescale}

Post-merger star formation can be rapidly quenched due to the fast gas consumption via intense starbursts and strong AGN feedback \citep[e.g.,][]{springel2005mergerquench, quai2021mergerquench, ellison2022mergerquench}. S0\,19 clearly shows the quick quenching processes following mergers at $z \sim 1.2$ ($\tau_{\rm Q} = 270$\,Myr) and $z \sim 0.33$ ($\tau_{\rm Q} = 15$\,Myr). Similarly, S0\,26 exhibits the effective central star formation prior to the first quenching at $z \sim 0.8$ ($\tau_{\rm Q} = 550$\,Myr). 

These S0 galaxies experience relatively short quenching timescales, with their enhanced SFR declining rapidly as most of the gas is depleted, compared to the longer timescales ($>$\,Gyr) typically observed in faded spiral scenarios \citep[e.g.,][]{peng2015strangulation, trussler2020starvation}. As shown in the previous section, enhanced AGN feedback may play a crucial role, such as driving gas outflows or gas heating \citep[e.g.,][]{di2005AGN, dubois2016thehorizon, zinger2020AGN}. We note that different definitions are used for quenching timescales \citep[e.g.,][]{wetzel2013tauQ, walters2022tauQ, park2023tauQ}, so one should be careful when comparing the exact values across different studies.

\subsection{Star-forming S0}
\label{sec: Discussion - Star-forming S0}

Based on ongoing central star formation and hints of galaxy interactions, such as misaligned kinematics, it has been argued that gas-rich minor mergers can rejuvenate S0s \citep[][]{rathore2022SFS0, xu2022SFS0}. In this study, we find that galaxies undergoing gas concentration acquire the smooth feature of an S0 earlier than the quenching of star formation  (see Figure~\ref{fig05_S019_mockimgs} at $z=1.3$ and Figure~\ref{fig09_S026_mockimgs} at $z=0.83$). This is because the shortage of outer disk gas suppresses late-type morphology while central star formation continues.

Figure~\ref{fig12_S021_gas_j} shows the case of S0\,21, the star-forming visually-S0 galaxy mentioned in Section~\ref{sec: Methods - Sample selection} (galaxy identified as $g$ in Figure~\ref{fig01_SFR_M}). It undergoes gas concentration along the increase in the mass of counter-rotating gas at $z\sim0.4$. It accretes floating gas from the vicinity without going through any mergers. As a result, the gas distribution is concentrated, and its morphology transforms to an S0, as shown in Figure~\ref{fig13_S021_mockimgs}. Due to the presence of central star formation, the galaxy is recognized as a star-forming S0. This provides a possible origin for centrally star-forming lenticular galaxies. Late-type galaxies that already have a lower gas fraction are more subject to this scenario, as even the small amount of gas accretion in retrograde orbit is able to remove the gas in the outskirts via hydrodynamic interaction \citep[][]{pizzella2004counter}.


\section{Summary and conclusions}
\label{sec: Summary and conclusions}

In this study, we used the \NH\, cosmological hydrodynamic simulation to investigate the formation pathways of S0s in field environments. 
We sampled two star-formation quenched, visually-lenticular galaxies of mass in the range of $10^{10-11}\,\msol$: S0\,19 and S0\,26. By tracing the evolution of physical properties and morphological changes, we found that each S0 follows a distinct formation pathway leading to S0 morphology and the quenching of star formation.

\vspace{+0.5cm}

(i) \emph{Merger: } As demonstrated by S0\,19, galaxy mergers drive the disk gas to the galaxy center, causing enhanced star formation and rapid galactic gas consumption.

(ii) \emph{Counter-rotating gas accretion: } As S0\,26 demonstrated, a galaxy may turn to lenticular through counter-rotating gas accretion from outside.
The hydrodynamic interaction between the gas components rotating in opposite or misaligned directions easily cancels pre-existing angular momentum information, driving gas inflow and central star formation and resulting in a lenticular morphology, in a similar manner as in mergers.

\vspace{+0.5cm}

We have found that gas angular momentum cancellation is an important mechanism in both merger and counter-rotating gas accretion scenarios. In both scenarios, quenching timescales are short ($<$\,Gyr).

We find star-forming, visually-lenticular galaxies as transient features in the course of becoming quenched S0 galaxies. They are undergoing the counter-rotating gas accretion pathway but with incomplete central gas depletion. This scenario may explain the central star formation and misaligned kinematics observed in star-forming S0s.

We emphasize that the counter-rotating gas accretion scenario could revise the previous interpretations regarding the origins of S0s. While mergers and retrograde gas accretion both lead to gas-star misaligned kinematics, gas accretion is a preferable mechanism for forming misaligned galaxies with a surviving stellar disk, as mergers tend to be more destructive \citep[][]{zhou2022misalinged}. Additionally, central gas migration and subsequent stellar growth may explain the large bulge components of S0s without relying solely on mergers \citep[][]{chen2016bulgegrowth}.

Despite the limited sample size, our study provides compelling evidence that both merger and counter-rotating gas accretion scenarios can account for the presence of S0 galaxies in field environments. Particularly, our work is the first cosmological simulation study to highlight counter-rotating gas accretion as a potential pathway for the formation of S0 galaxies. Future studies employing cosmological simulations that track gas dynamics and offer larger sample sizes are anticipated to validate this scenario further. 

\begin{acknowledgments}
\section*{Acknowledgements}

We thank San Han and Jinsu Rhee for numerous comments that improved the quality of the manuscript. 
This work was granted access to the HPC resources of CINES under the allocations c2016047637, A0020407637 and A0070402192 by Genci, KSC-2017-G2-0003, KSC-2020-CRE-0055 and KSC-2020-CRE-0280 by KISTI, and as a “Grand Challenge” project granted by GENCI on the AMD Rome extension of the Joliot Curie supercomputer at TGCC.
The large data transfer was supported by KREONET, which is managed and operated by KISTI.
We acknowledge the significant contribution of the radiative transfer code SKIRT, developed by Maarten Baes and Peter Camps. 
We thank the 8 researchers in the group who participated in the visual classification.
S.K.Y. acknowledges support from the Korean National Research Foundation (NRF-2020R1A2C3003769).
E.C. acknowledges support from the Korean National Research Foundation (RS-2023-00241934).
TK was supported by the National Research Foundation of Korea (NRF-2020R1C1C1007079). 
S.O. acknowledges support from the NRF grant funded by the Korean government (MSIT) (RS-2023-00214057). 
This study was funded in part by the NRF-2022R1A6A1A03053472 grant and the BK21Plus program.

\vspace{-0.5cm}
\end{acknowledgments}


\bibliography{ref}{}

%

\appendix
\restartappendixnumbering

\section{Morphology verification}
\label{sec: Appendix - Morphology verification}

Here, we present our classification result using a photometric approach to demonstrate that the visual stellar disks of NH galaxies are properly detected, which is essential for our S0 classification.  

\begin{figure}
\centering
\includegraphics[width=0.5\textwidth]{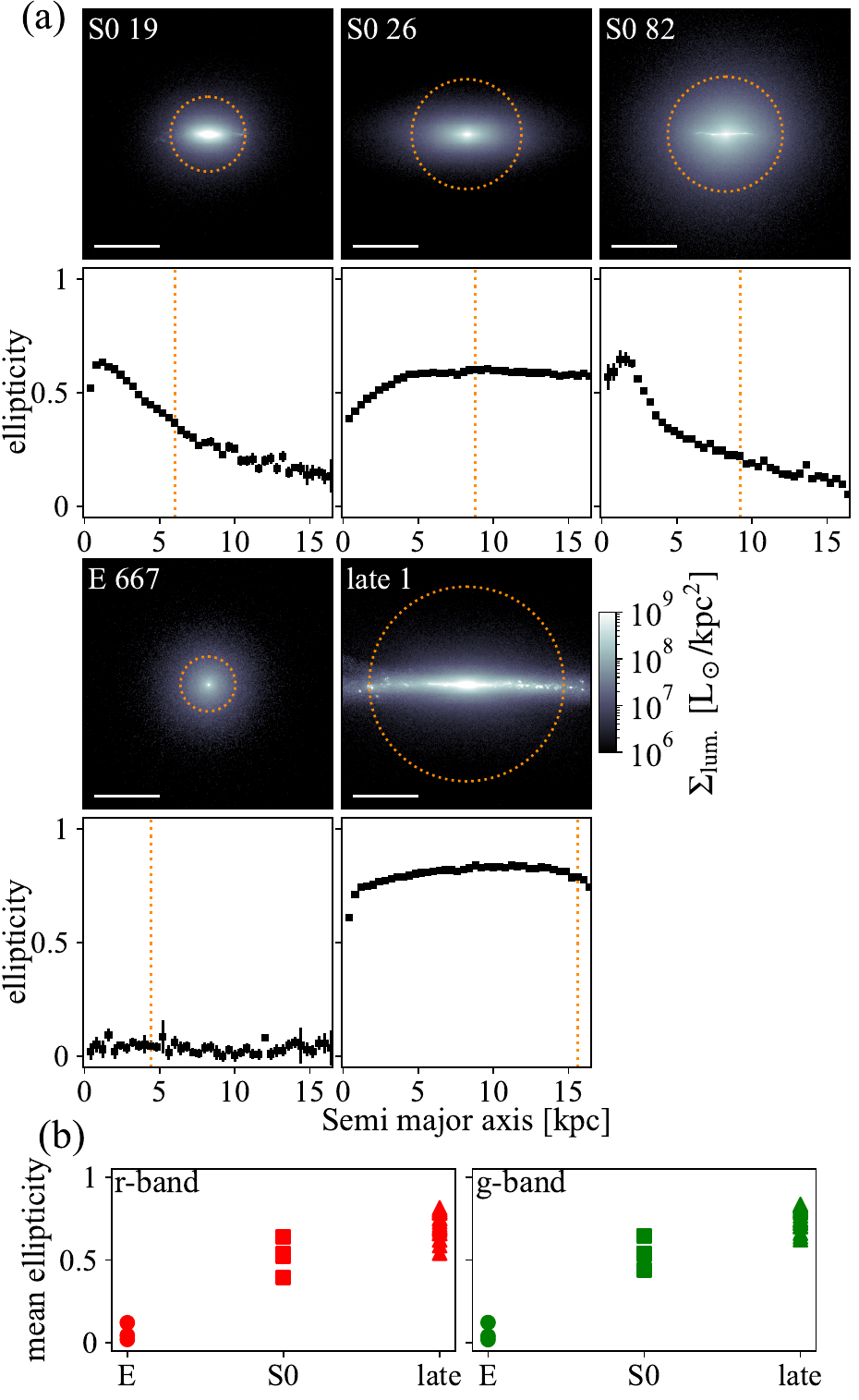}
\caption{Ellipticity radial profiles and comparison of mean ellipticities for Es, S0s, and late-type galaxies.
(a) shows the stellar surface luminosity maps and ellipticity radial profiles of five NH galaxies at $z=0.17$.
Each vertically aligned pair of a stellar map and a profile plot corresponds to the same galaxy.
The maps display the surface luminosity viewed edge-on in the r-band with 40\,pc-sized pixels.
The ellipticity profiles show the ellipticity of isophotes as a function of the semi-major axis with error bars.
The dotted orange lines in the plots represent where the surface luminosity of an isophote falls below $10^7\, \rm L_{\odot}\, kpc^{-2}$, marked as circles on the stellar maps correspondingly.
The white bars in the maps indicate 10\,kpc.
(b) compares the mean ellipticities of Es, S0s, and late-types measured in the r- (left) and g-band (right).
 }
\label{figA1_ell_profile}
\end{figure}

First, we measure ellipticities of elliptical isophotes from the stellar surface luminosity maps of Es, S0s, and late-types viewed edge-on at $z=0.17$, as shown in Figure~\ref{fig02_12gals_mockimgs}. We use the isophote package from photutils\footnote{\url{https://photutils.readthedocs.io/en/stable/isophote.html}} for ellipse fitting. Figure~\ref{figA1_ell_profile}(a) shows examples of the stellar luminosity maps and the ellipticities of isophotes as a function of their semi-major axis for each galaxy in the r-band. The ellipticity profiles vary with stellar distribution; for example, ellipticities at outer radii are lower for galaxies with a diffuse structure like S0\,19 and 82. However, Es constantly show low ellipticities across all radii with less diversity.

To obtain the mean ellipticity, we use isophotes with relatively high surface luminosity ($>10^7\, \rm L_{\odot}\,kpc^{-2}$), which possess small errors for fitting parameters. We then calculate the mean ellipticity of these isophotes. The mean ellipticities of Es, S0s, and late-types measured in the r- and g-bands are shown in Figure~\ref{figA1_ell_profile}(b). Regardless of band filters, there is a clear distinction between Es and S0s, with Es being more rounded and S0s flattened. Although the mean ellipticities of S0s and late-types overlap, late-types show higher values on average. Thus, we successfully differentiate S0s from Es by identifying visual stellar disks.

\begin{figure*}
\centering
\includegraphics[width=1\textwidth]{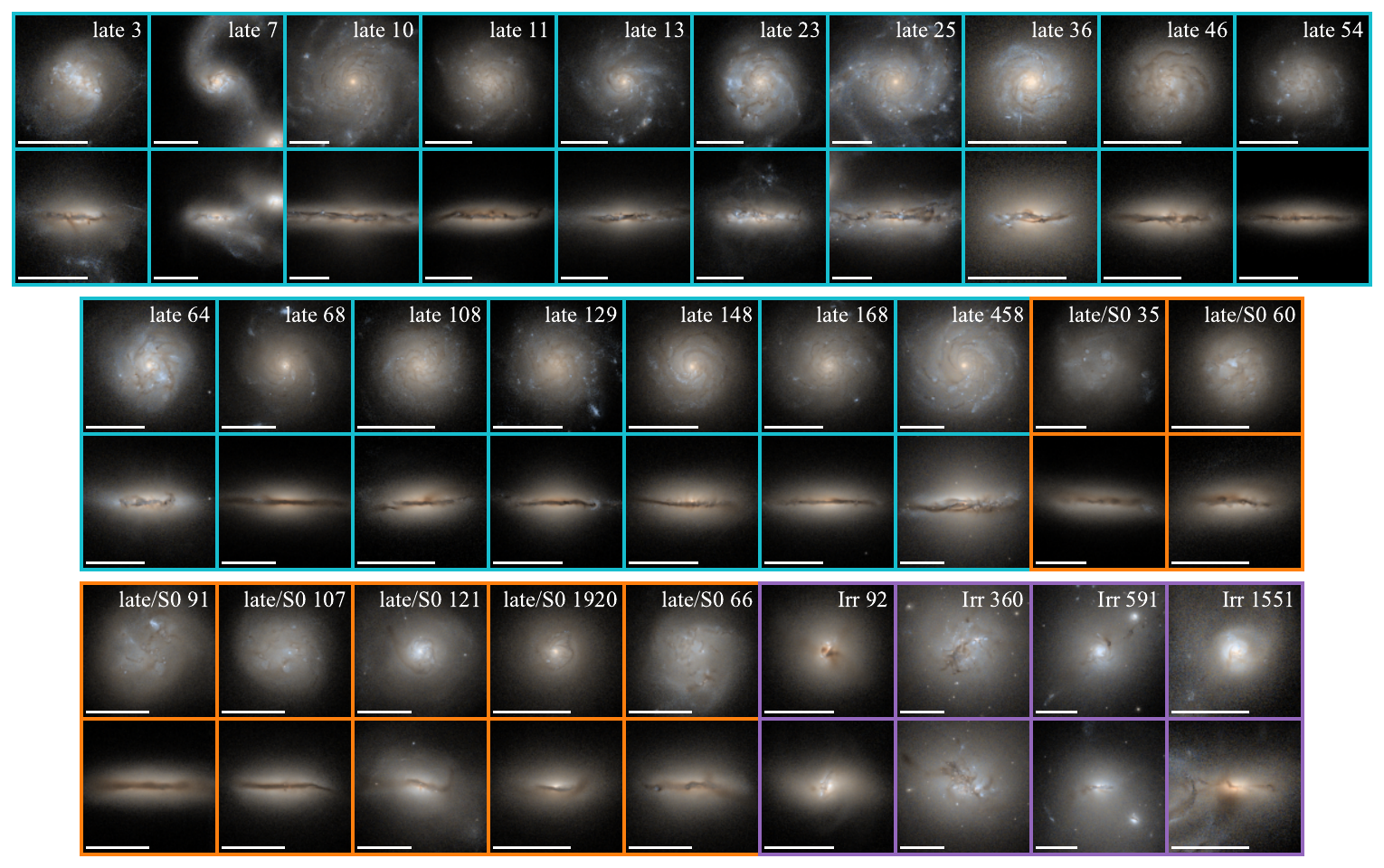}
\caption{Mock observation images of NH galaxies shown in face-on and edge-on views at $z = 0.17$.
This figure contains 28 galaxies with stellar masses between $10^{10}\,\msol$ and $10^{11}\,\msol$, which are not included in Figure~\ref{fig02_12gals_mockimgs}. 
For each galaxy, morphological type and ID are shown in the upper right corner of the face-on view panel. 
Panels are framed in different colors for different morphological types.
The white bars indicate 10\,kpc.
 }
\label{figA2_rest_mockimgs}
\end{figure*}

For readers interested in the complete results of our visual classification, Figure~\ref{figA2_rest_mockimgs} provides mock observation images of all other NH galaxies within the target stellar mass range, which are not shown in Figure~\ref{fig02_12gals_mockimgs}.

\section{Stellar kinematics}
\label{sec: Appendix - Stellar kinematics}

In this section, we examine the stellar kinematics of S0 galaxies to see how their visual morphology and kinematics relate.

We measure $V/\sigma_{\star}$, the rotation-to-dispersion ratio, to assess the stellar rotating motions of S0 galaxies using stellar particles within 2\re. The reference axis for this measurement is the vector sum of the angular momenta of stars. We obtain velocity components in cylindrical coordinates relative to this reference axis. $V/\sigma_{\star}$ is defined as the mean tangential velocity divided by the mean velocity dispersions of all stellar components, with mass weights considered. 

\begin{figure}
\centering
\includegraphics[width=0.5\textwidth]{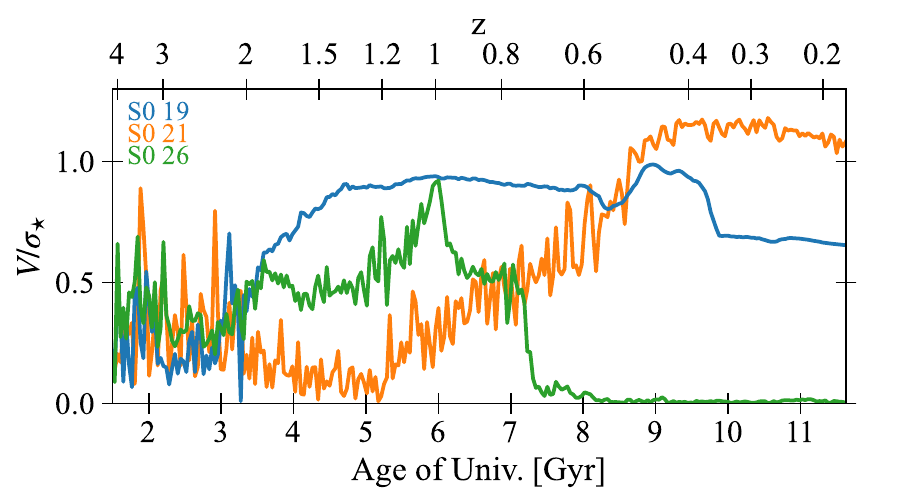}
\caption{Evolution of mass-weighted $V/\sigma_{\star}$ of S0s.
Each history is shown in different colors.
 }
\label{figB3_vsig}
\end{figure}

The histories of $V/\sigma_{\star}$ of S0s are shown in Figure~\ref{figB3_vsig}. The value is mostly affected by galaxy-galaxy interactions. For S0\,19, a mini merger triggers rapid star formation from counter-rotating gas around $z\sim0.4$, leading to the sudden decrease in $V/\sigma_{\star}$. 
In the case of S0\,26, $V/\sigma_{\star}$ reaches zero quickly after a major merger around $z=0.8$, as the remaining stellar disk is composed of both co- and counter-rotating disks that virtually cancel each other in the calculation of integrated $V/\sigma_{\star}$.

\end{document}